\documentclass[longauth]{aa}  
\usepackage{graphicx}
\usepackage{txfonts}
\usepackage{natbib,twoopt}
\usepackage{xcolor}
\usepackage{threeparttable}  
\usepackage{booktabs}
\usepackage{lscape}
\usepackage{multirow}

\bibpunct{(}{)}{;}{a}{}{,}           
\definecolor{xlinkcolor}{cmyk}{1,1,0,0}

\usepackage[bookmarks=true,
pdfnewwindow=true,
colorlinks=true,
linkcolor=xlinkcolor,
citecolor=xlinkcolor,
filecolor=xlinkcolor,
urlcolor=xlinkcolor,
final=true]{hyperref}

\def\Fig{\mbox{Figure~}}

\def\Tab{\mbox{Table~}}

\def\Sec{\mbox{Section~}}

\def\Eq{\mbox{Equation~}}

\def\App{\mbox{Appendix~}}

\newcommand{\rev}{}

\begin{document}

   \title{Going deeper into the dark with COSMOS-Web}
   
   \subtitle{JWST unveils the total contribution of Radio-Selected NIRfaint galaxies to the cosmic Star Formation Rate Density}

   \author{Fabrizio Gentile\thanks{\email{fabrizio.gentile@cea.fr}}
          \inst{1,2,3}
          \and
          Margherita Talia\inst{2,3}
          \and
          Andrea Enia\inst{2,3}
          \and
          Francesca Pozzi\inst{2,3}
          \and
          Alberto Traina\inst{3}
          \and
          Giovanni Zamorani\inst{3}
          \and
          Irham T. Andika\inst{4,5}
          \and
          Meriem Behiri\inst{6,7}
          \and
          Laia Barrufet\inst{8,9}
          \and
          Caitlin M. Casey\inst{10}
          \and
          Andrea Cimatti\inst{2,3}
          \and 
          Nicole E. Drakos\inst{11}
          \and
          Andreas L. Faisst\inst{12}
          \and
          Maximilien Franco\inst{1,10}
          \and
          Steven Gillman\inst{13,14}
          \and
          Marika Giulietti\inst{6,3}
          \and
          Rashmi Gottumukkala\inst{13}
          \and
          Christopher C. Hayward\inst{16}
          \and
          Olivier Ilbert\inst{17}
          \and
          Shuowen Jin\inst{13,14}
          \and
          Andrea Lapi\inst{6,18,19}
          \and
          Jed McKinney\inst{10}
          \and
          Marko Shuntov\inst{13,15}
          \and
          Mattia Vaccari\inst{20,21,7}
          \and
          Cristian Vignali\inst{2,3}
          \and
          Hollis B. Akins\inst{10}
          \and
          Natalie Allen\inst{13,15}
          \and
          Santosh Harish\inst{21}
          \and
          Henry Joy McCracken\inst{23}
          \and
          Jeyhan S. Kartaltepe\inst{22}
          \and
          Anton M. Koekemoer\inst{24}
          \and
          Daizhong Liu\inst{25}
          \and
          Louise Paquereau\inst{23}
          \and
          Jason Rhodes\inst{26}
          \and
          Micheal R. Rich\inst{27}
          \and
          Brant E. Robertson\inst{28}
          \and
          Sune Toft\inst{13,15}
          }

   \institute{CEA, IRFU, DAp, AIM, Université Paris-Saclay, Université Paris Cité, Sorbonne Paris Cité, CNRS, 91191 Gif-sur-Yvette, France
   \and
   University of Bologna, Department of Physics and Astronomy (DIFA), Via Gobetti 93/2, I-40129, Bologna, Italy
         \and
             INAF --- Osservatorio di Astrofisica e Scienza dello Spazio, via Gobetti 93/3 - 40129, Bologna - Italy
        \and
            Technical University of Munich, TUM School of Natural Sciences, Department of Physics, James-Franck-Str. 1, 85748 Garching, Germany
        \and
        Max-Planck-Institut für Astrophysik, Karl-Schwarzschild-Str. 1, 85748 Garching, Germany
        \and
        SISSA, Via Bonomea 265, I-34136 Trieste, Italy
        \and
        INAF/IRA, Istituto di Radioastronomia, Via Piero Gobetti 101, 40129 Bologna, Italy
        \and
        Department of Astronomy, University of Geneva, Chemin Pegasi 51, CH-1290 Versoix, Switzerland
        \and
        Institute for Astronomy, School of Physics \& Astronomy, University of Edinburgh, Royal Observatory, Edinburgh EH9 3HJ, UK
        \and
        The University of Texas at Austin, 2515 Speedway Boulevard Stop C1400, Austin, TX 78712, USA
        \and
        Department of Physics and Astronomy, University of Hawaii, Hilo, 200 W Kawili St, Hilo, HI 96720, USA
        \and
        Caltech/IPAC, MS314-6, 1200 E. California Blvd. Pasadena, CA, 91125, USA
        \and
        Cosmic Dawn Center (DAWN), Denmark
        \and
        DTU-Space, Technical University of Denmark, Elektrovej 327, DK-2800 Kgs. Lyngby, Denmark
        \and
        Niels Bohr Institute, University of Copenhagen, Jagtvej 128, DK-2200, Copenhagen, Denmark
        \and
        Center for Computational Astrophysics, Flatiron Institute, 162 Fifth Avenue, New York, NY 10010, USA
        \and
        Aix Marseille Univ, CNRS, CNES, LAM, Marseille, France
        \and
        IFPU - Institute for fundamental physics of the Universe, Via Beirut 2, 34014 Trieste, Italy
        \and
        INFN-Sezione di Trieste, via Valerio 2, 34127 Trieste,  Italy
        \and
        Inter-University Institute for Data Intensive Astronomy, Department of Astronomy, University of Cape Town, 7701 Rondebosch, Cape Town, South Africa
        \and
        Inter-University Institute for Data Intensive Astronomy, Department of Physics and Astronomy, University of the Western Cape, 7535 Bellville, Cape Town, South Africa
        \and
        Laboratory for Multiwavelength Astrophysics, School of Physics and Astronomy, Rochester Institute of Technology, 84 Lomb Memorial Drive, Rochester, NY 14623, USA
        \and
        Institut d’Astrophysique de Paris, UMR 7095, CNRS, and Sorbonne Université, 98 bis boulevard Arago, F-75014 Paris, France
        \and
        Space Telescope Science Institute, 3700 San Martin Drive, Baltimore, MD 21218, USA
        \and
        Purple Mountain Observatory, Chinese Academy of Sciences, 10 Yuanhua Road, Nanjing 210023, China
        \and
        Jet Propulsion Laboratory, California Institute of Technology, 4800 Oak Grove Drive, Pasadena, CA 91001, USA
        \and
        Department of Physics and Astronomy, UCLA, PAB 430 Portola Plaza, Box 951547, Los Angeles, CA 90095-1547
        \and
        Department of Astronomy and Astrophysics, University of California, Santa Cruz, 1156 High Street, Santa Cruz, CA 95064, USA}

   \date{Received ??; Accepted ??}

\abstract{We present the first follow-up with JWST of radio-selected NIRfaint galaxies as part of the COSMOS-Web survey. By selecting galaxies detected at radio frequencies ($S_{\rm 3 GHz}>11.5$ $\mu$Jy; i.e. S/N$>5$) and with faint counterparts at NIR wavelengths (F150W$>26.1$ mag), we collect a sample of 127 likely dusty star-forming galaxies (DSFGs). We estimate their physical properties through SED fitting, compute the first radio luminosity function for these types of sources, and their contribution to the total cosmic star formation rate density. Our analysis confirms that these sources represent a population of highly dust-obscured ($\langle A_{\rm v} \rangle \sim3.5$ mag), massive ($\langle M_\star \rangle \sim10^{10.8}$ M$_\odot$) and star-forming galaxies ($\langle {\rm SFR} \rangle\sim300$ M$_\odot$ yr$^{-1}$) located at $\langle z \rangle\sim3.6$, representing the high-redshift tail of the full distribution of radio sources. Our results also indicate that these galaxies could dominate the bright end of the radio luminosity function and reach a total contribution to the cosmic star formation rate density equal to that estimated only considering NIR-bright sources at $z\sim4.5$. Finally, our analysis further confirms that the radio selection can be employed to collect statistically significant samples of DSFGs, representing a complementary alternative to the other selections based on JWST colors or detection at FIR/(sub)mm wavelengths.}


   \maketitle

\section{Introduction}

Understanding how galaxies evolved from the Big Bang to current time is one of the main open questions of modern astrophysics. The answer is commonly thought to reside in deep galaxy surveys, collecting objects at different cosmic times and potentially unveiling their evolutionary links. Most of these surveys are performed at optical and near-IR (NIR) wavelengths. Therefore --- at $z>3$ --- these surveys sample a spectral range more affected by dust extinction (see e.g. \citealt{Salim_20} for a review). For this reason, a population of dust-obscured galaxies is constantly missed by these surveys, potentially biasing all the results achieved with them \citep[e.g.][]{Simpson_14,Franco_18,wang_19,talia_21,enia_22,Behiri_23,Barrufet_23,Gottumukkala_23,Gentile_24,Williams_24}. These dusty star-forming galaxies (DSFGs) lacking a counterpart at optical/NIR wavelengths have assumed several names: HST-dark, H-dropout, NIR-dark, OIR-dark, and NIRfaint galaxies. All these classifications are commonly based on the lack of counterpart in the optical/NIR regimes coupled with a bright flux at longer wavelengths, potentially tracing ongoing star formation. These tracers can be in the mid-IR (MIR; see e.g. \citealt{wang_19,Barrufet_23,Gottumukkala_23,Williams_24}), (sub)mm (e.g. \citealt{Simpson_14,Franco_18,gruppioni_20,Smail_21}), or radio (e.g. \citealt{Chapman_01,talia_21,enia_22,Behiri_23,VanDerVlugt_23,Gentile_24}). {\rev Regardless of the selection, several} studies targeting these sources agree that they represent a population of massive ($M_\star>10^{10}$ M$_\odot$) and dust-obscured ($A_{\rm v}>3$ mag) galaxies mainly located at $z\sim2-3$ and beyond. However, a consensus is still missing about their actual contribution to the cosmic star formation rate density (SFRD; i.e. the average amount of stellar mass formed in the Universe each year and in each cubic Mpc, see e.g. the results by \citealt{wang_19,talia_21,enia_22,Barrufet_23,Behiri_23,VanDerVlugt_23,Williams_24}), the overlap between the different populations (e.g. \citealt{Smail_02,talia_21,McKinney_24,Williams_24}), and their relationship with the broader population of galaxies (e.g. \citealt{Cochrane_24}).

The advent of the James Webb Space Telescope (JWST) gives us the opportunity to study these populations on a common ground. Its unprecedented sensitivity allows us to detect --- for the first time --- the rest-frame optical radiation from these sources, reducing the uncertainties on their photometric redshifts and stellar masses (see e.g. \citealt{Barrufet_23,Gottumukkala_23}) and --- more broadly --- to fairly compare the different populations (see e.g. \citealt{perez-gonzalez_23,Gillman_24,Hodge_24,McKinney_24}).

In this paper, we focus on the radio-selected NIRfaint galaxies \citep{talia_21}. These sources are defined as radio-detected sources without a counterpart in the optical/NIR regimes at the depths commonly reached by ground-based facilities (25-27 mag; see e.g. \citealt{weaver_22}). Given radio is a dust-unbiased tracer of star formation (see e.g. \citealt{kennicutt_12}), these sources are excellent candidates as dusty star-forming galaxies (DSFGs; see e.g. \citealt{Chapman_01,talia_21,enia_22,Behiri_23,Gentile_24}). 

{\rev Compared with other possible methods to collect DSFGs, the radio selection presents some advantages. Firstly, the higher resolution and sensitivity of radio interferometers than previous generation FIR/sub(mm) facilities (with beam sizes larger than 10"; see e.g. \citealt{Swinyard_10,Dempsey_13}) such as \textit{Herschel} or the SCUBA-2 camera equipped on the James Clerk Maxwell Telescope (JCMT; see e.g. the initial studies in this field by \citealt{Smail_97,Hughes_98,Burgarella_13,Gruppioni_13} and \citealt{casey_14} for a review). These properties allow the easier association of multi-wavelength counterparts and the discovery of fainter sources. Moreover, the higher mapping speed of radio interferometers with respect to modern facilities such as the Atacama Large Millimetre Array (ALMA) or the NOrtern Extended Millimetre Array (NOEMA) makes it possible to explore larger cosmic volumes and --- therefore --- obtain larger samples less affected by cosmic variance and poor statistics. Secondly, the radio selection is expected to be more robust to the contamination by red and passive galaxies affecting the MIR selection  (see e.g. the fraction of non-dusty sources reported in the MIR-selected samples by \citealt{wang_19,perez-gonzalez_23,Barrufet_24})}. 

The downside of such selection is represented, on the one hand, by the positive $k$-correction at radio frequencies, limiting our possibilities to select very high-\textit{z} DSFGs that are more easily detected at (sub)mm wavelengths (thanks to the negative $k$-correction in that regime; see e.g. \citealt{Casey_21}). On the other hand, the radio selection needs to take into accounts the possible contribution by Active Galactic Nuclei (AGN; also able to emit at radio frequencies, see e.g. the review by \citealt{Tadhunter_16}). For doing so, we focus on the sources detected in the COSMOS field: one of the most famous extra-galactic fields in modern astronomy (see e.g. \citealt{Scoville_07,Koekemoer_07}) with a good photometric coverage from radio to X-rays allowing us to estimate the AGN contribution to our sample of targets with the help of ancillary data. We also take advantage of the new JWST coverage of the COSMOS field granted by the COSMOS-Web survey \citep{Casey_23}.

The main goals of this work consist in the estimation of the physical properties of these sources, their first radio luminosity function (never estimated before for NIR-dark/faint galaxies) and --- consequently --- their contribution to the cosmic SFRD. These results will then be employed to compare our targets with other notable populations of optically/NIRfaint galaxies in the current literature.

The paper is structured as follows. In \Sec\ref{sec:data}, we describe the data and the sample selection. In \Sec\ref{sec:bagpipes}, we estimate the physical properties of our targets through SED fitting and the possible AGN contribution. In \Sec\ref{sec:LF} we estimate the luminosity function of our sources and in \Sec\ref{sec:SFRD} we use this result to compute how much our galaxies contribute to the cosmic SFRD. We discuss our results in \Sec\ref{sec:discussion} and --- finally --- we draw our conclusions in \Sec\ref{sec:summary}.

Throughout this paper, we assume a standard flat $\Lambda$CDM cosmology with the parameters $[h,\Omega_{\rm M},\Omega_\Lambda]=[0.7,0.3,0.7]$. We also assume a \citet{Chabrier_03} initial mass function and the AB photometric system \citep{Oke_83}.

\section{Data}
\label{sec:data}
\subsection{JWST photometry}
\label{sec:JWST}

The JWST photometry for our sources comes from the COSMOS-Web survey (GO \#1727, PIs Kartaltepe \& Casey; \citealt{Casey_23}), a cycle 1 treasury program consisting in the NIRCam and MIRI imaging of the COSMOS field. The program includes a contiguous NIRCam mosaic covering the central region of the field ($\sim$0.54 deg$^2$) in the four filters F115W, F150W (the short-wavelength filters; SW hereafter), F277W, and F444W (long-wavelength; LW). In parallel with the NIRCam mosaic, COSMOS-Web also includes a MIRI mosaic covering a total (non-contiguous) area of 0.19 deg$^2$ in the F770W filter. A full description of the COSMOS-Web program can be found in \citet{Casey_23}, while the data reduction procedure is described in detail in M. Franco et al., (\textit{in prep.}). 

In this work, we use the NIRCam and MIRI photometry extracted with \texttt{SourceXtractor++} (\texttt{SE++}; \citealt{Bertin_20,Kummel_20}). This software performs the detection on a positive-truncated $\chi^2$ combination of the four NIRCam filters PSF-homogenized to F444W. Each source is then modeled with a Sérsic profile \citep{Sersic_63}. The best-fitting model is then convolved with the PSFs of the five (NIRCam and MIRI; {\rev PSF FWHM=0.04-0.14'' and 0.27'', respectively}) scientific maps in order to extract the fluxes and the related uncertainties. {\rev The full catalog includes photometry for more than 800,000 sources, down to a $5\sigma$ limiting magnitude of $m=28.0$ in F444W}. A full description of the procedure followed to build the COSMOS-Web catalogs can be found in M. Shuntov, L. Paquereau et al., (\textit{in prep.}).

\subsection{Radio data}
\label{sec:radio_data}

The radio data analysed in this paper come from the VLA-COSMOS Large Program \citep{smolcic_17}: a radio survey performed with the \textit{Karl Jansky Very Large Array} (VLA) covering the full COSMOS field ($\sim2$ deg$^2$) at a frequency of 3 GHz. The survey reaches a quite uniform rms of $\sigma=2.3$ $\mu$Jy beam$^{-1}$ and a spatial resolution of 0.75''. To perform our sample selection, we cross-match the photometric catalog of COSMOS-Web with the public catalog by \citet{smolcic_17} with a matching radius of 0.7'' (as in \citealt{Gentile_24}). This procedure gives in output a sample of 3196 galaxies. However, given the better spatial resolution of NIRCam than that achieved with the VLA, 139 radio sources have multiple NIRCam galaxies falling in the matching radius. Given the impossibility to recognize the actual object originating the radio emission, we exclude these sources from the final sample. 

\subsection{Sample selection}
\label{sec:selection}

We perform a sample selection resembling those previously employed by \citet{talia_21}, \citet{Behiri_23}, and \citet{Gentile_24} to collect their sample of ``radio-selected NIRdark galaxies" in the COSMOS field, but taking advantage of the new JWST photometry coming from the COSMOS-Web survey. 

We start from the parent sample assembled in \Sec\ref{sec:radio_data} and we select our galaxies through the following criteria:

\begin{equation}
\centering
\begin{cases}
        S_{\rm 3GHz} & >11.5 \,\, \mu{\rm Jy}\\
        F150W & >26.1 \,\, {\rm mag}
\end{cases}
\end{equation}

Clearly --- by construction --- all the galaxies in the final sample also need to be detected in the detection image employed to assemble the COMSOS-Web catalog. 

The first criterion requires that the sources are robustly ($S/N>5$) detected at radio frequencies, while the latter requires that they would not be detected in the NIR filter H ($\lambda\sim1.6$ $\mu$m) at the $3\sigma$ depth of the COSMOS2020 catalog (2'' aperture; see \citealt{weaver_22}). The full sample includes 127 galaxies in the 0.54 deg$^2$ observed in the COSMOS-Web survey (55 with MIRI coverage, $\sim32\%$ of the full sample, 44 with a robust detection at S/N$>$3).

Compared with the previous criteria employed by \citet{talia_21}, \citet{Behiri_23}, and \citet{Gentile_24}, we include in our sample sources with lower values of the S/N  {\rev at 3 GHz}. The previous studies, indeed, employed a $S/N>5.5$ cut to reduce the number of possible spurious sources. However, the availability of a strong prior such as the NIRCam imaging allows us to relax this criterion, potentially including in our sample some higher-\textit{z} sources previously missed due to their faintness at radio frequencies (e.g. the spectroscopically-confirmed source at $z\sim5$ reported by \citealt{Jin_19}, with a reported S/N at 3 GHz of 5.2). 

Moreover, those studies required the lack of counterpart in the COSMOS2020 catalog, whose detection was performed on a $\chi^2-$image including the four VISTA filters $Y, J, H$, and $Ks$ and the two optical filters $i$ and $z$ from HSC \citep{weaver_22}. However, the low resolution of these images did not completely allow a counterpart-matching for some optically-bright galaxies blended with nearby sources (see the discussion in \citealt{Gentile_24}). With these improved criteria and with the high-resolution of the NIRCam maps, we aim to select a less contaminated sample of ``NIRdark" galaxies. Since these sources are now detected at the new depths reached by NIRCam, in the following we will refer to our galaxies as radio-selected NIRfaint (RS-NIRfaint).

\subsection{Ancillary data}
\label{sec:ancillary}

Our targets are located in the COSMOS field. Therefore, an almost complete photometric coverage is available for them. We include in our analysis the following data:

\begin{itemize}
    \item \textbf{Optical-to-MIR}: The \texttt{SE++} software employed to extract the NIRCam and MIRI photometry can also be applied to other ancillary data, by fitting each source with the parametric model computed on the NIRCam maps (once convolved with the PSF of low-resolution maps). These additional data include:
    \begin{enumerate}
        \item The optical maps obtained during the Subaru Strategic Program (SSP DR3; \citealt{Aihara_19}) performed with the HyperSupreme Cam (HSC) mounted on the Subaru telescope and those obtained with the Advanced Camera for Surveys (ACS) equipped on HST \citep{Koekemoer_07}.
        \item The NIR data obtained during the UltraVISTA survey (DR6; \citealt{McCracken_12}) performed with the VIRCAM instrument of the VISTA telescope.
        \item The MIRI data obtained as part of the Cosmic Dawn Survey of the COSMOS field \citep{Moneti_22} performed with the Infrared Array Camera (IRAC) equipped on the \textit{Spitzer} Space Telescope.
    \end{enumerate}
    
    The footprint of all these surveys overlap with the full COSMOS-Web area. In order to cover wavelength ranges not included in our NIRCam and MIRI photometry, we include in our catalog the data in the eight filters $g,r,i,z,y, F814W, Ks$ and in the first channel of IRAC ($\lambda\sim3.8$ $\mu$m). The depths and additional details on the employed maps can be found in \citet{weaver_22} and in M. Shuntov, L. Paquereau et al., (\textit{in prep.}).
    
    \item \textbf{FIR/(sub)mm:} We cross-match our catalog with the most updated version of the super-deblended catalog (\citealt{jin_18}; Jin et al., \textit{in prep.}), including deblended photometry from the PACS and SPIRE instruments equipped on the \textit{Herschel} space telescope that observed the COSMOS field during the surveys described in \citet{Lutz_11} and \citet{Oliver_12}. From the same catalog, we also retrieve photometry at 870 $\mu$m from the deblending of the SCUBA-2 maps obtained during the S2COSMOS survey by \citet{Simpson_19}. Since the super-deblended employs the 3 GHz sources of the VLA-COSMOS survey as priors, all our galaxies have an entry in that catalog. More in detail, 72 sources in our sample have at least one detection at S/N$>$3 in at least one PACS or SPIRE filter. Similarly, 31 objects are detected at S/N$>$3 in the 870 $\mu$m maps. All the other objects have upper limits. We also obtain ALMA photometry in the (sub)mm range for 38 of our sources ($\sim 30\%$ of the sample) by cross-matching with the most recent catalog (v20220606) from the Automated mining of the public ALMA Archive in the COSMOS field (A3COSMOS; \citealt{Liu_19,Adscheid_24}).
    \item \textbf{Radio:} Finally, we retrieve data at radio frequencies (3, 1.4, and 1.28 GHz) by cross-matching with the public catalogs from the VLA-COSMOS large program \citep{smolcic_17,Schinnerer_07} and the early data release of the MIGHTEE survey performed with MeerKAT \citep{Jarvis_16}. Given the lower resolution of the MIGHTEE maps ($\sim8''$), we only consider the 1.28 GHz radio fluxes of 84 isolated sources (i.e. without another 3 GHz object within 8''). 
\end{itemize}

\begin{figure*}
    \centering
    \includegraphics[width=0.5\textwidth]{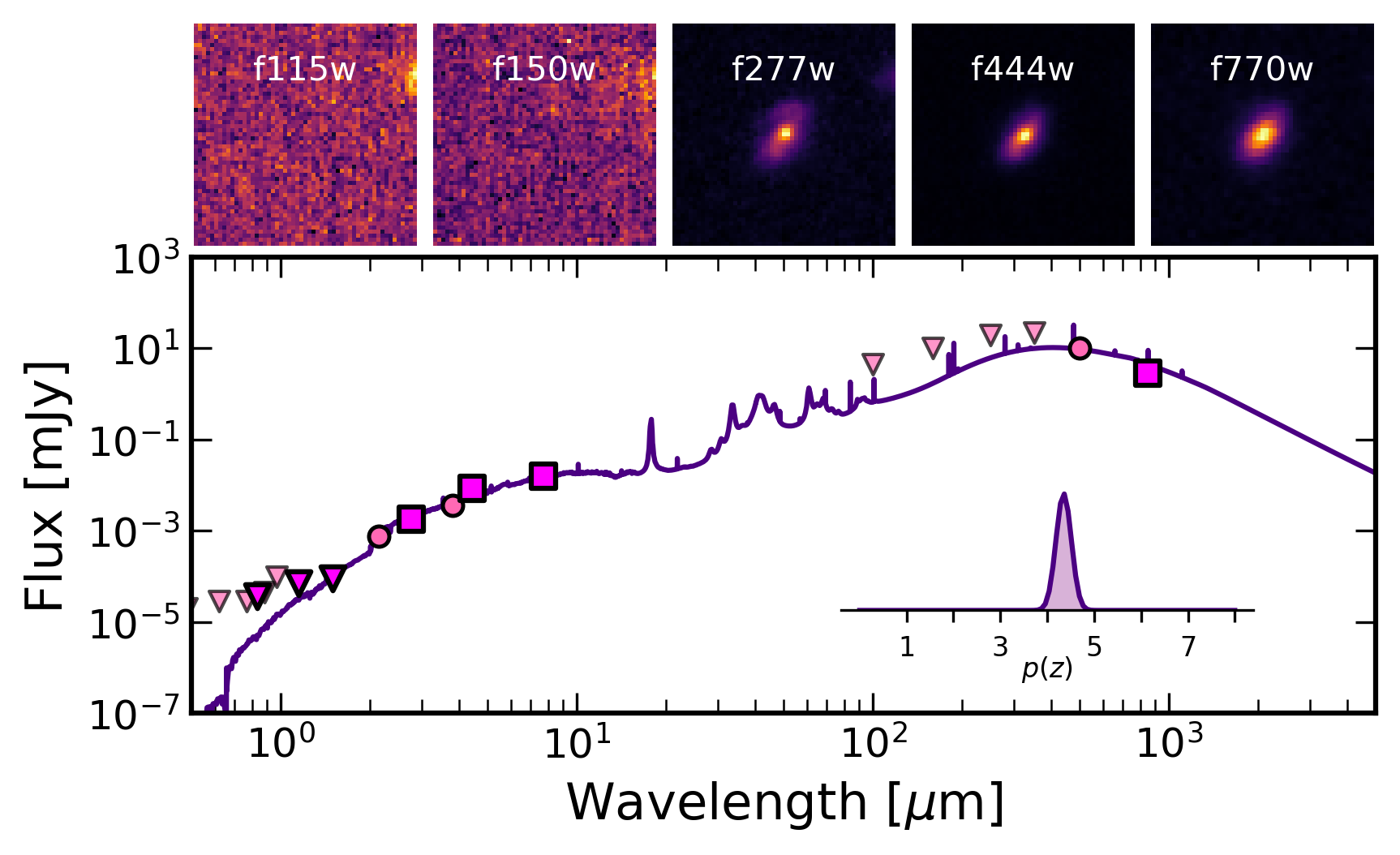}\includegraphics[width=0.5\textwidth]{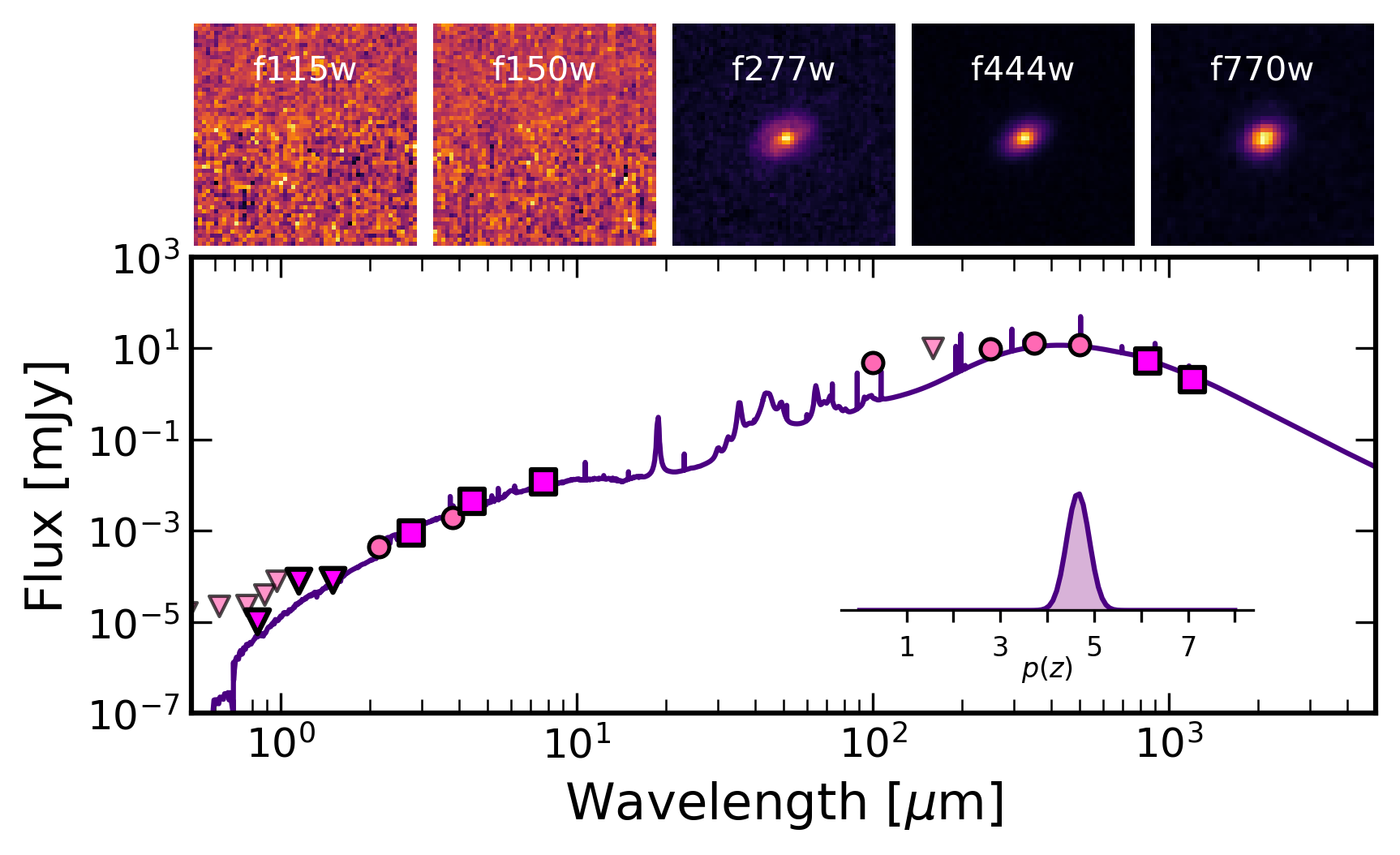}
    \\\includegraphics[width=0.5\textwidth]{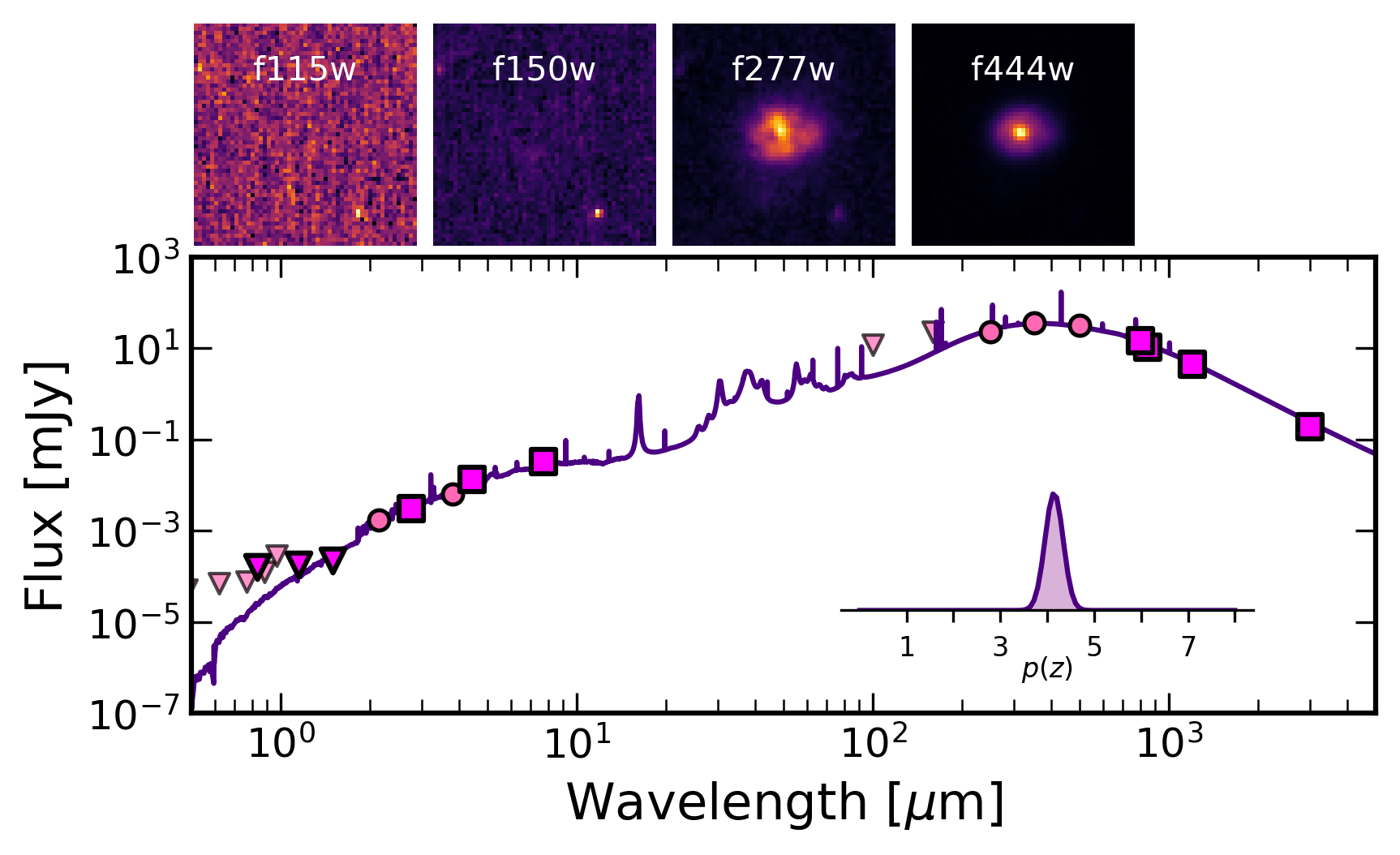}\includegraphics[width=0.5\textwidth]{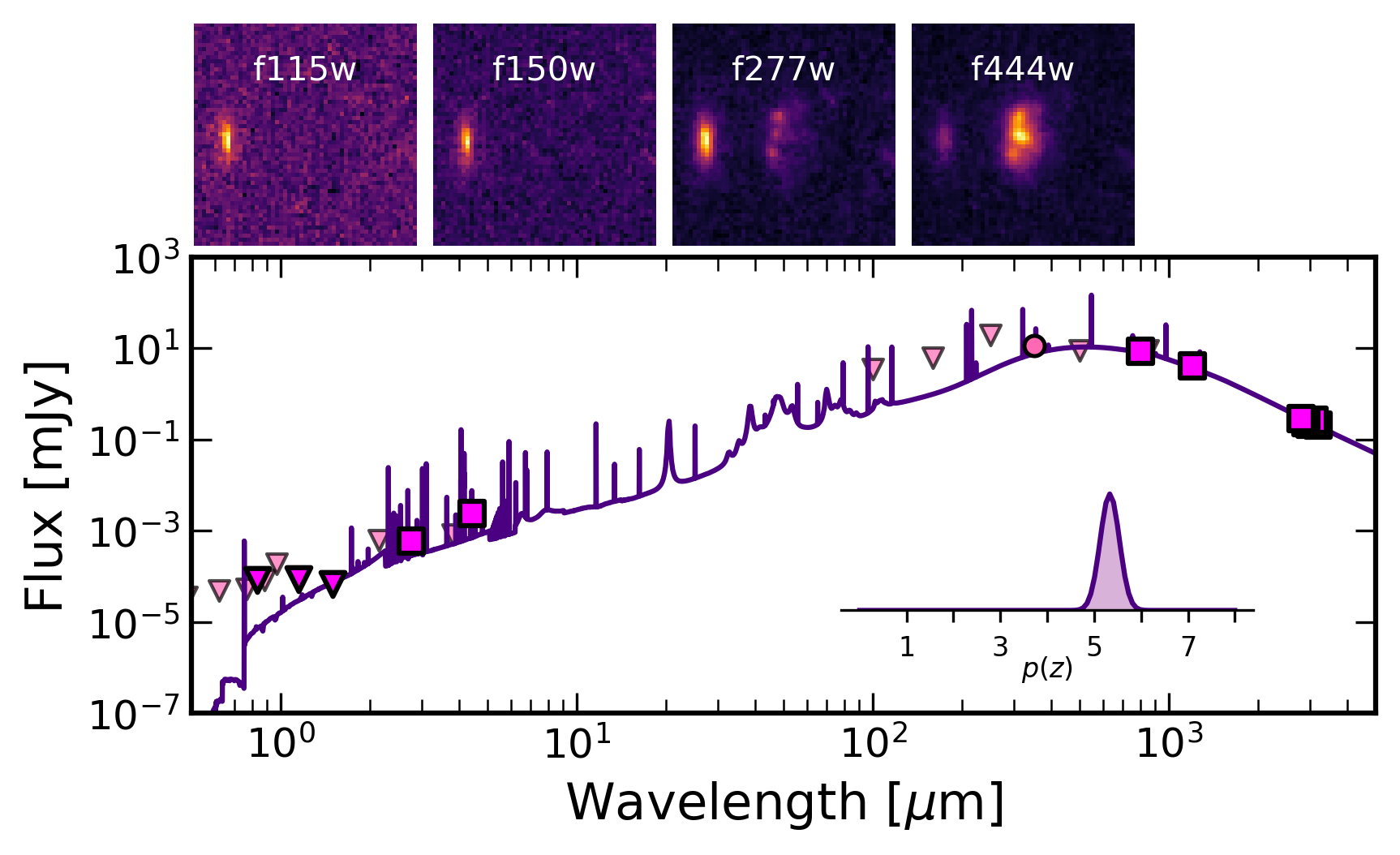}
    \caption{Some examples of the SED fitting performed with \texttt{Cigale} on the galaxies included in the highest redshift bin. The squares indicate the photometric points from JWST, A3COSMOS and the 870$\mu$m from the deblending of the S2COSMOS maps by \citet{Simpson_19}. The circular points report the ancillary photometry from HSC, VISTA, Spitzer, and from the super-deblending of the Herschel Maps. The inset shows the Gaussianized $p(z)$ computed with \texttt{Cigale}. The top row shows the cutouts (3 arcsec side) in the NIRCam and MIRI filters of COSMOS-Web.}
    \label{fig:SEDs}
\end{figure*}

\section{SED fitting}
\label{sec:bagpipes}

Our SED fitting is performed with the ``Code Investigating GALaxy Emission" (\texttt{Cigale}; \citealt{boquien_19}), a software based on the principle of energy balance between the dust attenuation and its thermal emission at longer wavelengths. The input catalog includes all the photometry available for our sources with the only exception of the radio data (employed later to estimate the AGN fraction in our sources; see \Sec\ref{sec:AGN}). Our setup includes the stellar populations by \citet{Bruzual_03}, combined through a delayed exponentially-declining star formation history with an additional recent burst of star formation. The stellar emission is extincted through a \citet{Charlot_00} attenuation law, while the dust emission is included in our templates through the models by \citet{Draine_14}. Finally, the nebular emission is accounted for through a series of models computed with \texttt{Cloudy} \citep{Ferland_13}. The resulting templates are redshifted on a grid spanning the range $z\in[0,8]$ with a step of 0.05. The full list of parameters employed in our SED fitting is included in the \App\ref{App:SEDFitting}. The good convergence of the SED-fitting procedure is ensured by the distribution of the reduced $\chi^2$, with a median value of $\langle \chi_{\nu}^2 \rangle=1.1$ and $95\%$ of the sample with a $\chi_{\nu}^2<5$. Some examples of the SED-fitting output are reported in \Fig\ref{fig:SEDs}.

\subsection{Photometric redshifts}
\label{sec:photoz}

\begin{figure}
    \centering
    \includegraphics[width=\columnwidth]{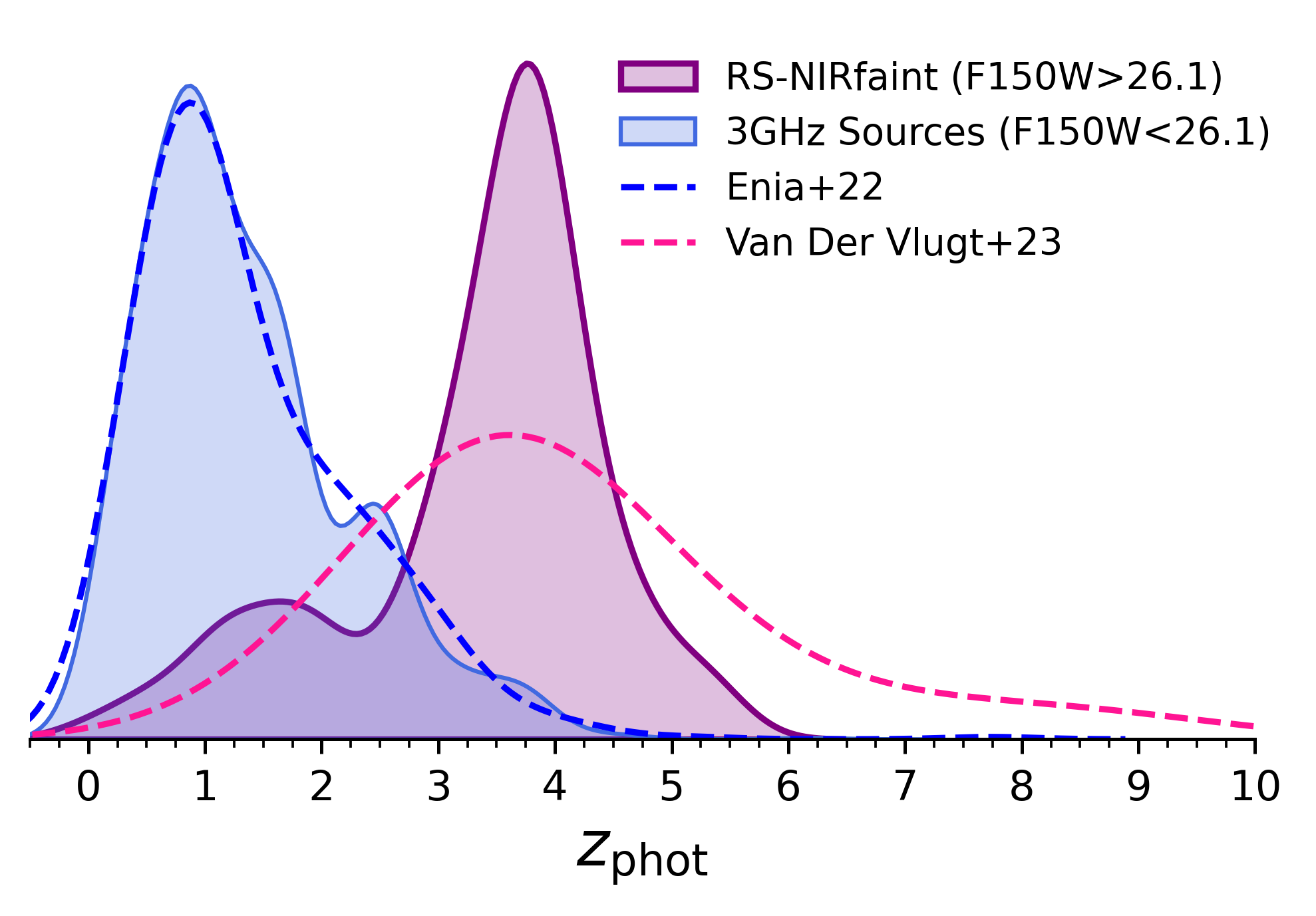}
    \caption{Distribution of the photometric redshifts of our RS-NIRfaint galaxies in COSMOS-Web. Our sources are reported as the purple solid line, while the complementary sample of 3 GHz objects with F150W$>$26.1 mag is reported as the blue solid line. For reference, we also show the photo-\textit{z} computed by \citet{enia_22} on their sample of radio sources (with optical/NIR counterparts) in the GOODS-N field (blue dashed line) and those estimated by \citet{VanDerVlugt_23} for their sample of optically/NIR-faint galaxies in their deeper COSMOS-XS survey (dashed pink line).}
    \label{fig:photozs}
\end{figure}

The distribution of the photometric redshifts for our sources is reported in \Fig\ref{fig:photozs} {\rev (the other samples showed for reference are described in detail in \Sec\ref{sec:other_pops})}. The distribution is quite peaked at $\langle z \rangle \sim3.6$, with a $1\sigma$ dispersion (given as the symmetrized interval between the 16th and the 84th percentile) of 0.8. The availability of the new NIRCam and MIRI photometry allows us to sample the rest-frame optical/NIR emission of our galaxies, reducing the uncertainty on the photometric redshifts. The median $\delta z/(1+z)$ for our galaxies
is 0.08, nearly half of what we would obtain by removing the JWST photometry from the SED fitting performed with \texttt{Cigale} (0.15). The spectroscopic coverage of our sample is not sufficient to allow a proper testing of our photometric redshift. However, thanks to the collection of spectroscopic redshift in COSMOS (A. A. Khostovan et al., \textit{in prep.}), we found four sources in our sample with a spec-\textit{z}. These are ID20010161 ($z_{\rm spec}=5.051$) from \citet{Jin_19} (see also \citealt{Gentile_24c}) and AS2COS0002 ($z_{\rm spec}=4.600$), AS2COS0011 ($z_{\rm spec}=4.783$), and  AS2COS0014 ($z_{\rm spec}=2.920$) from \citet{Chen_22}. These spectroscopic redshifts are well recovered by our SED fitting, with the first three objects having a discrepancy lower than $2\sigma$ and only the last one having a spec-\textit{z} at 5$\sigma$ from the photo-\textit{z}. The median value (computed on this small sample) of the $|\Delta z|/(1+z)$ is 0.09.

\begin{figure}
    \centering
    \includegraphics[width=\columnwidth]{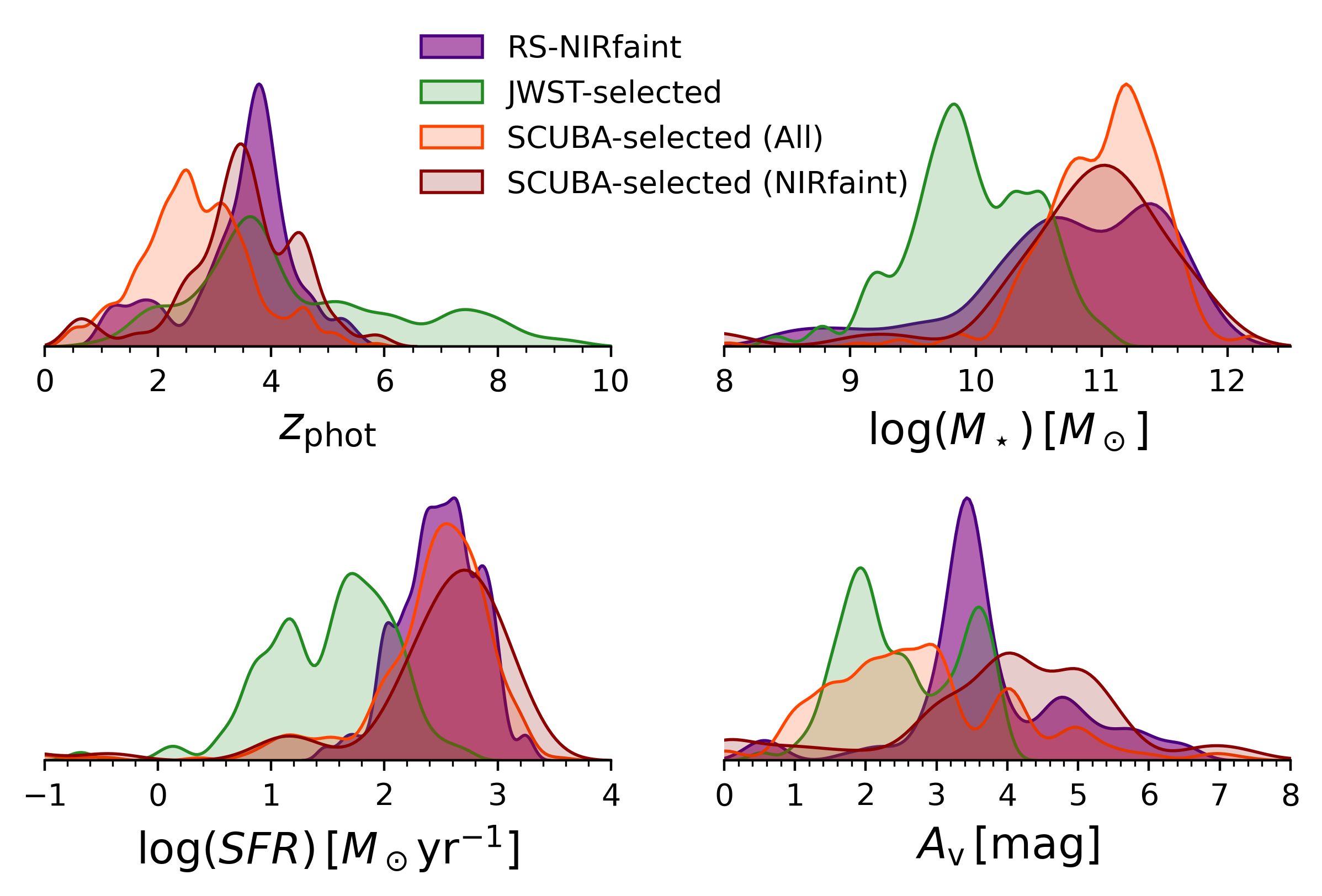}
    \caption{Main properties (photometric redshift, stellar mass, SFR, and dust extinction) of our RS-NIRfaint galaxies. We show --- for comparison --- the same properties computed by \citet{Gottumukkala_23} and \citet{McKinney_24} for their sample of JWST-selected and SCUBA-selected DSFGs as the green and orange solid lines, respectively. To allow a fair comparison, we only include the sources by \citet{Gottumukkala_23} with $F150W>26.1$ mag and exclude the sources flagged as possible AGN.}
    \label{fig:properties}
\end{figure}

\subsection{Physical properties}
\label{sec:properties}

With \texttt{Cigale}, we also estimate the stellar masses and dust attenuation for our galaxies. Since we do not have any constraints on the rest-frame UV SED of our sources, we do not use the SFR computed through SED fitting for our analysis. Instead, we compute the SFR from the radio luminosity --- after accounting for the possible AGN contribution --- as described in \Sec\ref{sec:AGN}\footnote{{\rev The same lack of constraints in the rest-UV affects, in theory, also the estimation of $A_{\rm v}$. Nevertheless, since we have some constrains from the rest-optical and from the employment of an SED-fitting code relying on the energy balance, we decide to still report our estimate of $A_{\rm v}$ in \Tab\ref{tab:properties}}}. The results of this procedure are summarized in \Tab\ref{tab:properties} and in \Fig\ref{fig:properties}. These distributions picture the RS-NIRfaint galaxies as a population of highly dust-obscured ($\langle A_{\rm v} \rangle \sim3.5$ mag), massive ($\langle M_\star \rangle \sim10^{10.8}$ M$_\odot$) and star-forming galaxies ($\langle {\rm SFR} \rangle\sim300$ M$_\odot$ yr$^{-1}$) located at $\langle z \rangle\sim3.6$. \Fig\ref{fig:properties} also shows for reference the same properties computed for other notable populations of dusty star-forming galaxies (i.e. those selected with JWST by \citealt{Gottumukkala_23} and with SCUBA/ALMA by \citealt{McKinney_24}). A full comparison between the RS-NIRfaint galaxies and these other samples will be the focus of \Sec\ref{sec:other_pops}.

\begin{table}
\centering
\begin{threeparttable}
\renewcommand{\arraystretch}{1.1}
\caption{Median properties of our RS-NIRfaint galaxies}
\label{tab:properties}
\begin{tabular}{cccc}
\toprule
Property & Median & $\sigma$ & Unit \\
\midrule
$z_{\rm phot}$ & 3.6 & 0.8 & -\\
$\log(M_\star)$ & 10.8 & 0.7 & M$_\odot$\\
$\log$(SFR) & 2.5 & 0.4 & M$_\odot$ yr$^{-1}$\\
$A_{\rm v}$ & 3.5 & 0.9 & mag\\
\bottomrule
\end{tabular}
\textbf{Note:}\\
The $\sigma$ of each distribution is reported as the symmetrized interval between the 84th and the 16th percentiles.
\end{threeparttable}
\end{table}

\subsection{AGN contribution}
\label{sec:AGN}

Since we are dealing with galaxies selected at radio frequencies, we need to consider that part of that emission could be due to nuclear activity and not to star formation. For doing so, we follow the procedure established by \citet{Ceraj_18} to measure the so-called ``AGN fraction" ($f_{\rm AGN}$), quantifying the AGN contribution to the radio luminosity. 

We start by computing the infrared luminosity ($L_{\rm IR}$) of our sources by integrating the best-fitting SED given in output by \texttt{Cigale} in the range [8,1000] $\mu$m. The $L_{\rm IR}$ is then employed to compute the $q_{\rm TIR}$ parameter (quantifying the ratio between infrared and radio emission) for our galaxies as:
\begin{equation}
    q_{\rm TIR}= \log\left(\frac{L_{\rm IR} [W]}{3.75\times10^{12} {\rm Hz}}\right)-\log\left(\frac{L_{\rm 1.4 GHz,tot}}{\rm W Hz^{-1}}\right)
\end{equation}
where $L_{\rm 1.4,tot}$ is the radio luminosity at 1.4 GHz computed from the radio flux at 3 GHz. The conversion between these quantities is done by computing the radio slope between 1.4 (or 1.28) GHz and 3 GHz (for the galaxies with a measured flux at these frequencies, see \Sec\ref{sec:ancillary}) or by assuming a fixed slope of $\alpha=-0.7$ (commonly employed for star-forming galaxies, see e.g. \citealt{novak_17}).

The $q_{\rm TIR}$ of star-forming galaxies is found to correlate with redshift. A possible parametrization is that found by \citet{delhaize_17}\footnote{We underline that the analysis described in \citet{delhaize_17} is based on the same radio survey employed to select the galaxies in this study.} as:
\begin{equation}
\label{eq:Delhaize}
    q_{\rm TIR} (z) = (2.88\pm0.03)(1+z)^{-0.19\pm0.01}
\end{equation}
We can take advantage of this relation to estimate the AGN fraction as 
\begin{equation}
    f_{\rm AGN}(q,z)=10^{q-\bar{q}(z)}
\end{equation}
where $q$ is the $q_{\rm TIR}$ measured for our galaxy and $\bar{q}(z)$ is the same quantity expected for a galaxy at the same redshift following the relation by \citet{delhaize_17}. Since in that study the $q_{\rm TIR}(z)$ relation is found to have an intrinsic scatter of 0.26, we assign a $f_{\rm AGN}=0$ to all the sources with $q-\bar{q}<0.26$ (i.e. those compatible within $1\sigma$ with the relation expected for star-forming galaxies).

\begin{figure}
    \centering
    \includegraphics[width=\columnwidth]{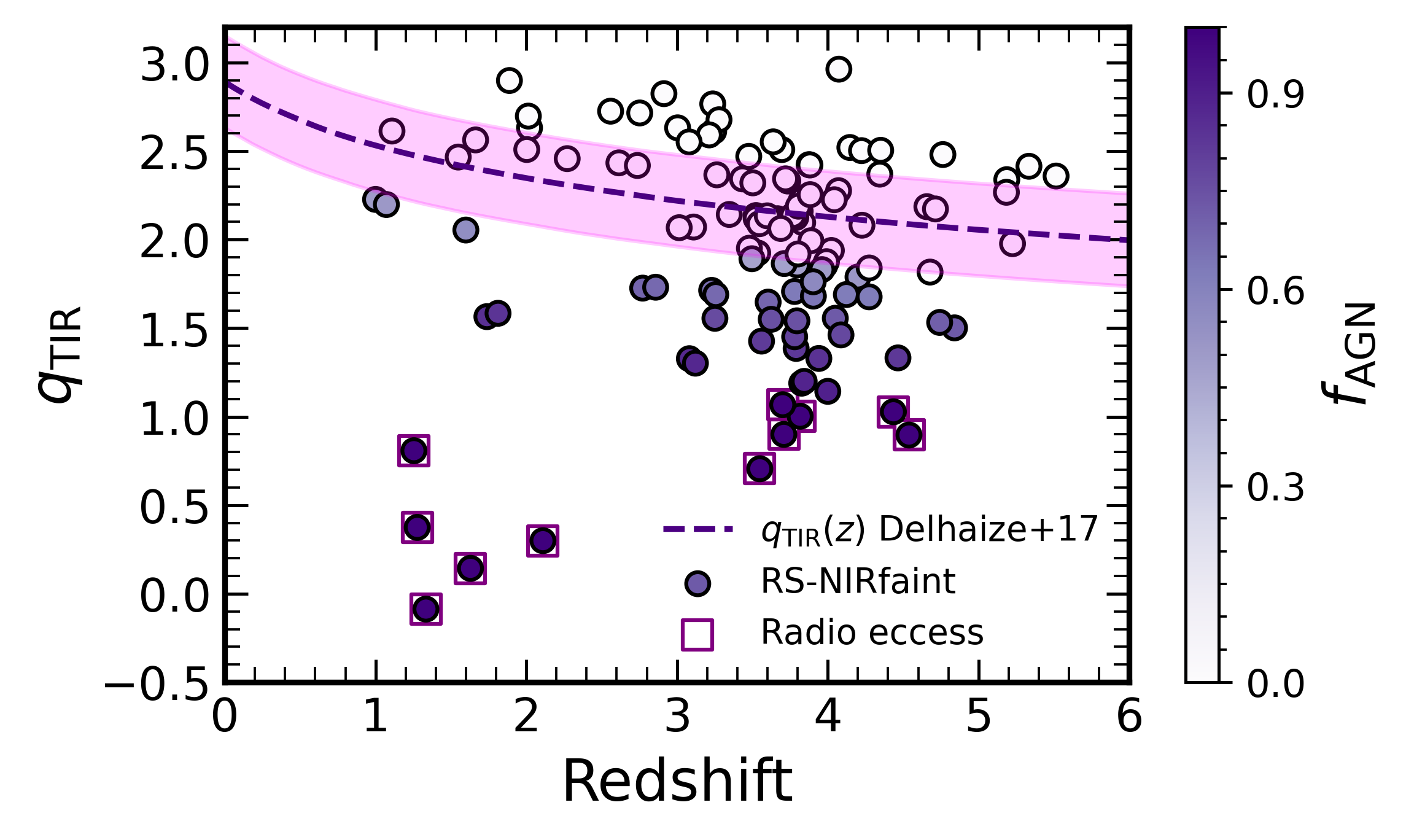}
    \caption{$q_{\rm TIR}$ parameter for our galaxies as a function of the redshift. The dashed line reports the $q_{\rm TIR}(z)$ relation by \citet{delhaize_17}, with the shaded area indicating the intrinsic scatter of the relation. Our galaxies are color-coded for their AGN fraction, measuring the contribution of nuclear activity to the radio luminosity. Galaxies classified as ``radio-excess" (i.e. with $f_{\rm AGN}>0.9$) are surrounded by a red square and removed from the sample.}
    \label{fig:qTIR}
\end{figure}

As visible in \Fig\ref{fig:qTIR}, most of the galaxies in our sample have low values of the AGN fraction ($f_{\rm AGN}<0.3$) and just 13 sources ($\sim 10\%$ of the full sample) are dominated by the AGN emission ($f_{\rm AGN}>0.9$; see e.g. \citealt{enia_22}) and --- therefore --- are removed from the sample. For all the other sources, we can account for the AGN contribution by computing the radio luminosity due to star formation as
\begin{equation}
    L_{\rm 1.4 GHz}=L_{\rm1.4 GHz, tot} (1-f_{\rm AGN})
\end{equation}

This value is then used to compute the SFR following \citet{novak_17}:
\begin{equation}
\label{eq:Novak}
    {\rm SFR}[M_\odot \, {\rm yr}^{-1}]=10^{-24} 10^{q(z)} \frac{L_{\rm 1.4 GHz}}{\rm W \, Hz^{-1}}
\end{equation}

by assuming the same $q_{\rm TIR}(z)$ relation found by \citet{delhaize_17} and presented in \Eq\ref{eq:Delhaize}. {\rev As an additional test, we cross-match our sample with the two publicly-available X-ray catalogs of the COSMOS field (from the C-COSMOS and COSMOS Legacy surveys: \citealt{Elvis_09,Civano_16}) using as matching radius the positional uncertainties reported in there. We find a single source (cid\_1076) with a counterpart in our sample. Given the relatively shallow depth of the X-ray observations in COSMOS, a detected source should have an X-ray luminosity higher than $L_{\rm x}>10^{42} \ {\rm erg \ s^{-1}}$ (more precisely, $L_{\rm x}\sim10^{44} \ {\rm erg \ s^{-1}}$ in the range 2-10 keV assuming the photo-\textit{z} by \texttt{Cigale} of $z\sim3.5$), likely hosting an un-obscured AGN. For this reason, we do not include this source in the rest of our analysis.}

\section{Luminosity Function}
\label{sec:LF}

We compute the luminosity function of our RS-NIRfaint galaxies by dividing our sample in three equally-populated redshift bins, covering the ranges $2.5-3.3$, $3.3-3.8$, and $3.8-5.5$. In each redshift bin, we divide the galaxies in several bins of (AGN-corrected) radio luminosity with a fixed width of 0.35 dex. To better sample the range of luminosities we employ bins that are overlapping by half of their width. In each combined redshift-luminosity bin, we compute the luminosity function following the $1/V_{\rm max}$ method \citep{Schmidt_68} as:
\begin{equation}
    \Phi(L,z)=\frac{1}{\Delta \log L}\sum_i\frac{1}{V_{\rm max,i}}
\end{equation}
where $\Delta\log L$ is the log-amplitude of the luminosity bin and the sum extends to all the galaxies in that bin. For each galaxy, $V_{\rm max}$ is computed as 
\begin{equation}
\label{eq:Vmax}
    V_{\rm max}=\sum_{z=z_{\rm min}}^{z_{\rm max}}[V(z+\Delta z)- V(z)]\, C(z)    
\end{equation}
where we employ a fixed width of the redshift shells of $\Delta z=0.05$. The $C(z)$ term is the completeness function and it can be written as the product of two terms:
\begin{equation}
    C(z)=C_{\rm A}C_{\rm 3 GHz}(z)
\end{equation}
The first one accounts for the limited area covered by the COSMOS-Web survey. In our case it can be written as the ratio between the area of the survey and that of the whole sky:
\begin{equation}
    C_{\rm A}=\frac{0.54 \, \, {\rm deg}^2}{41253 \, \,  {\rm deg}^2}
\end{equation}
The second term accounts for the incompleteness of the radio survey where our galaxies are selected. The completeness function of the 3 GHz survey performed during the VLA-COSMOS large program is reported in \citet{smolcic_17} for resolved and un-resolved sources. Since in our sample we have both kinds of sources, for each flux we consider the average between the two values. 

The sum in \Eq\ref{eq:Vmax} extends between two values $z_{\rm min}$ and $z_{\rm max}$ indicating the minimum and maximum redshift in which our target would be included in our sample according to our selection criteria. As described in \Sec\ref{sec:selection}, our galaxies need to have a radio S/N$>5$ at 3 GHz and be faint in the F150W filter of NIRCam\footnote{According to our selection, our targets also need to have a counterpart in the COSMOS-Web catalogs. However, since all our galaxies are robustly detected in the LW filters of NIRCam, we neglect this additional constraint.}.

\begin{figure}
    \centering
    \includegraphics[width=\columnwidth]{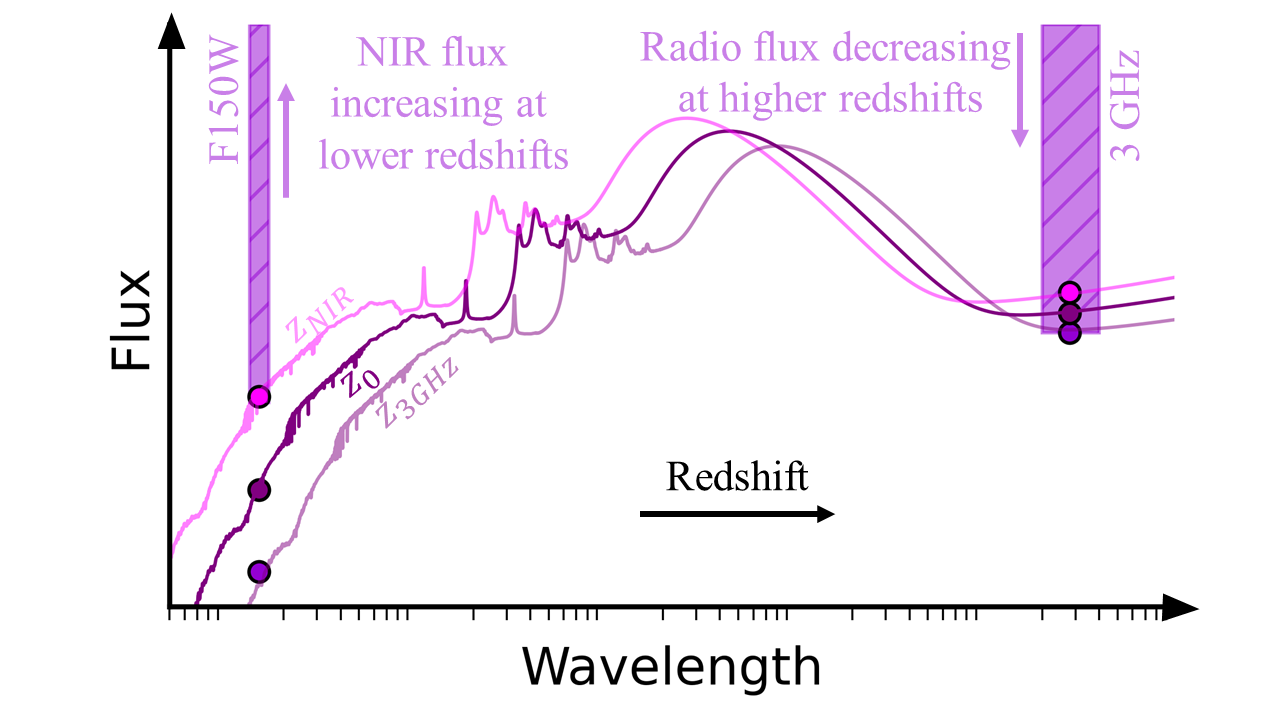}
    \caption{Sketch showing the physical meaning of the $z_{\rm min}$ and $z_{\rm max}$ values employed in the estimation of the LF. These values account for the redshift range in which each target could be found according to our selection criteria. See the definitions of $z_{\rm NIR}$ and $z_{\rm 3 GHz}$ in \Sec\ref{sec:LF}}.
    \label{fig:ZMinMax}
\end{figure}

As shown in \Fig\ref{fig:ZMinMax}, these properties are affected by the redshift of each object: at higher redshifts, our galaxy becomes fainter in the radio, up to a certain $z_{\rm3 GHz}$ where it becomes too faint to satisfy our S/N cut. Similarly, moving from high to low redshift, the F150W filter starts to sample less dust-obscured region of the SED, until a certain $z_{\rm NIR}$ where the galaxy becomes too bright to satisfy our magnitude cut ($F150W>26.1$ mag). Both these effects are accounted for in the definition of $z_{\rm min}$ and $z_{\rm max}$ as:

\begin{equation}
\centering
\begin{cases}
    z_{\rm min}&={\rm max}\,(z_{\rm min, bin}, z_{\rm NIR}) \\
    z_{\rm max}&={\rm min}\,(z_{\rm max, bin}, z_{\rm 3 GHz})
\end{cases}
\end{equation}

where $z_{\rm min, bin}$ and $z_{\rm max, bin}$ indicate the lower and upper limits of the considered redshift bin. The values $z_{\rm NIR}$ and $z_{\rm 3 GHz}$ are estimated for each galaxy by redshifting the best-fitting rest-frame SED on a grid with a step of $\Delta z=0.01$ and measuring the expected fluxes at 3 GHz and in F150W filter of NIRCam at each redshift.

\begin{figure*}
    \centering
    \includegraphics[width=\textwidth]{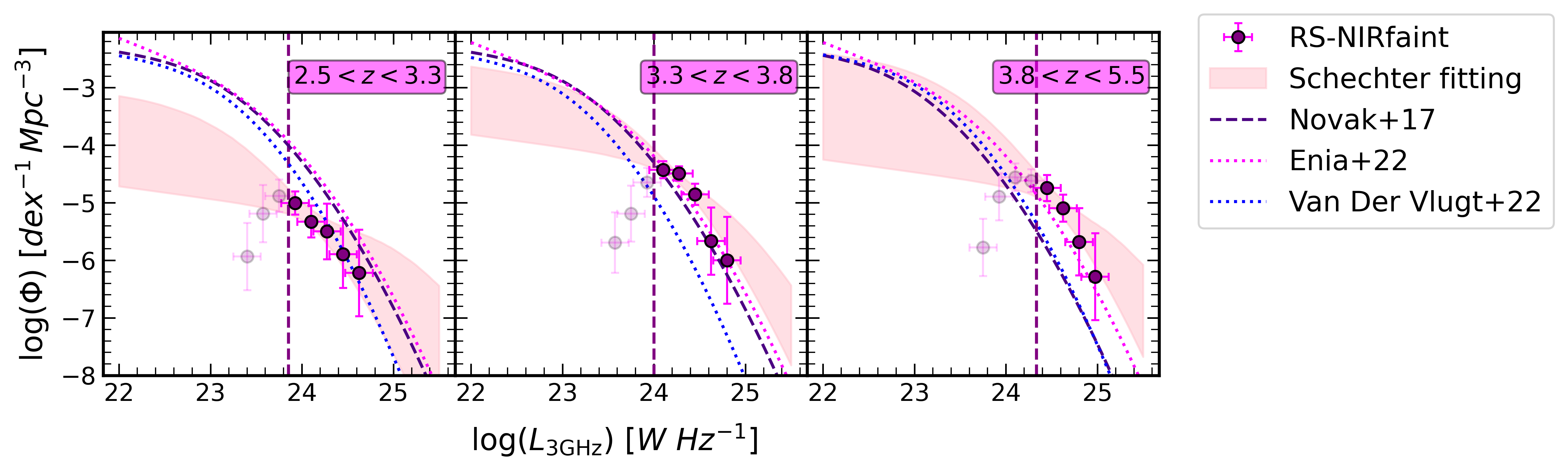}
    \caption{Radio luminosity function at 3 GHz of our RS-NIRfaint galaxies. The points show the values computed as described in \Sec\ref{sec:LF}, while the dashed vertical lines indicate the minimum radio luminosity observable in a given redshift bin given the sensitivity of our survey. All the points corresponding to fainter luminosities (shaded points) are not included in the fitting. The shaded areas report the fitting of the modified Schechter function, with its 1 $\sigma$ uncertainty. For reference, we also report the radio luminosity function (converted to 3 GHz by assuming a standard radio slope of $\alpha=-0.7$) estimated by \citet{novak_17}, \citet{enia_22}, and \citet{VanDerVlugt_22} on their samples of radio NIR-bright galaxies.}
    \label{fig:LF}
\end{figure*}

To account for the uncertainties in the photo-\textit{z} and radio fluxes at 3 GHz of our sources, we perform a Monte-Carlo integration by extracting 5000 random values from the probability distributions of these quantities. For the redshifts, we directly sample the Gaussianized $p(z)$ computed by \texttt{Cigale}, while for the radio fluxes we extract values from a Gaussian distribution centered on the mean value of the radio flux of each source and standard deviation equal to the related uncertainty. The value of the LF in each bin of redshift and radio luminosity is computed as the median value of all the iterations. Similarly, we assume as the upper and lower uncertainties the 16th and 84th percentiles, respectively. To account for the likely underestimation of the uncertainties in the less populated bins --- due to the contribution of Poissonian uncertainties affecting low number counts --- we correct these values in the bins with less than five galaxies with the $1\sigma$ confidence intervals reported in \citet{Gehrels_86}. Finally, we underline that our uncertainties do not include any contribution from cosmic variance. The values of the radio LF and the relative uncertainties are shown in \Fig\ref{fig:LF}.

\subsection{Modified Schechter function}
\label{sec:Schecther}

In each redshift bin, we fit the LF of our sources with a modified Schechter function \citep{Saunders_90}, a common choice for radio and (sub)mm luminosity functions (see e.g. \citealt{novak_17,gruppioni_20,enia_22,VanDerVlugt_22,Traina_24});
\begin{equation}
    \Phi(L)=\Phi_\star\left(\frac{L}{L_\star}\right)^{1-\alpha}{\rm exp}\left[-\frac{1}{2\sigma^2}\log^2\left(1+\frac{L}{L_\star}\right)\right]
\end{equation}
Due to the limited number of data-points available for fitting and the lack of constraints on the faint-end of the LF, we fix the two slope factors ($\alpha$ and $\sigma$) to the values found by \citet{novak_17} in their analysis of the sample of radio sources (with optical/NIR counterparts) in the VLA-COSMOS large program. These values are $\sigma=0.63$ and $\alpha=1.22$. We underline that the latter value is also in agreement with that found by \citet{VanDerVlugt_22} in the deeper COSMOS-XS survey, better sampling the faint end of the radio LF. Possible caveats related to this choice are discussed in \Sec\ref{sec:caveats}.

We fit the two remaining free parameters ($\Phi_\star$ and $L_\star$, giving the normalization and the ``knee" of the LF, respectively) with a Monte-Carlo Markov Chain (MCMC) performed with the Python library \texttt{emcee} \citep{Foreman_emcee}. We adopt a flat prior on the two parameters: $\log(\Phi_\star)\in[-6,-2]$ Mpc$^{-3}$ dex$^{-1}$ and $\log(L_\star)\in[20,26]$ W Hz$^{-1}$. We only include in the fitting procedure the points with $L>L_{\rm min}$, where $L_{\rm min}$ represents the minimum radio luminosity observable in a redshift bin given the $5\sigma$ sensitivity of our radio survey. Moreover, since the LF is computed in overlapping bins of radio luminosity, we only include in the fitting half of the data-points to obtain un-correlated uncertainties.

The best-fitting parameters of the modified Schechter function --- for each redshift bin --- are reported in \Tab\ref{tab:LF}, while the fitted function (with its $1\sigma$ confidence interval) is reported as the shaded pink area in \Fig\ref{fig:LF}.

\begin{table*}
\centering
\begin{threeparttable}
\renewcommand{\arraystretch}{1.1}
\caption{Best-fitting parameters of the modified Schechter function fitted to our radio luminosity function at 3 GHz. As a reference, we report the same values obtained by \citet{enia_22} in the GOODS-N field considering NIR-bright galaxies. Their luminosities (originally at 1.4 GHz) are converted assuming a standard radio slope $\alpha=-0.7$.}
\label{tab:LF}
\begin{tabular}{ccccc}
\toprule
 & \multicolumn{2}{c}{This work} & \multicolumn{2}{c}{Enia+22} \\ 
$z$ & $\log(\Phi_\star)$ & $\log(L_\star)$ & $\log(\Phi_\star)$ & $\log(L_\star)$ \\
 & [Mpc$^{-3}$ dex$^{-1}$] & [W Hz$^{-1}$] & [Mpc$^{-3}$ dex$^{-1}$] & [W Hz$^{-1}$]\\
\midrule
$2.5<z<3.3$ & $-4.4^{+0.8}_{-1.1}$ & $23.1^{+0.7}_{-1.2}$ & $-2.6^{+0.5}_{-0.5}$ & $22.7^{+0.3}_{-0.3}$\\
$3.3<z<3.8$ & $-3.5^{+0.6}_{-0.8}$ & $23.1^{+0.4}_{-0.6}$ & \multirow{2}{*}{$-2.7^{+0.4}_{-0.5}$} &  \multirow{2}{*}{ $22.8^{+0.3}_{-0.2}$}\\
$3.8<z<5.5$ & $-3.6^{+1.1}_{-1.0}$ & $23.2^{+0.5}_{-1.0}$ & &\\
\bottomrule
\end{tabular}
\end{threeparttable}
\end{table*}

\begin{table}
\centering
\begin{threeparttable}
\renewcommand{\arraystretch}{1.1}
\caption{Contribution of the RS-NIRfaint galaxies to the cosmic SFRD. These values are obtained by integrating the LF in the range of radio luminosity covered by our observations (``observed") and in the full range of luminosities (``extrapolated").}
\label{tab:SFRD}
\begin{tabular}{ccc}
\toprule
$z$ & \multicolumn{2}{c}{$\log(\rho_{\rm SFR})$} \\
& \multicolumn{2}{c}{[$\times 10^{-3}$ M$_\odot$ yr$^{-1}$ Mpc$^{-3}$]}\\
& (Observed) & (Extrapolated)\\
\midrule
$2.5<z<3.3$ & $1.3^{+0.7}_{-0.4}$ & $6^{+14}_{-3}$  \\
$3.3<z<3.8$ & $8^{+2}_{-2}$  & $30^{+34}_{-9}$ \\
$3.8<z<5.5$ & $4^{+2}_{-1}$  & $30^{+58}_{-17}$  \\
\bottomrule
\end{tabular}
\end{threeparttable}
\end{table}

\section{Contribution to the cosmic SFRD}
\label{sec:SFRD}

In each redshift bin, we compute the contribution of the RS-NIRfaint galaxies to the total cosmic SFRD by integrating the analytic expression of the modified Schechter function weighted for the SFR related to each radio luminosity:

\begin{equation}
    {\rm SFRD}(z)=\int_{L_{\rm min}}^{L_{\rm max}} \Phi(L,z)\, {\rm SFR}(L,z) \, d\log L	
\end{equation}\\
where ${\rm SFR}(L,z)$ is the expression reported in \Eq\ref{eq:Novak} and $z$ is the mean redshift of each bin. In this integral, we consider the full posterior distribution of the parameters in the modified Schechter function obtained from the MCMC in \Sec\ref{sec:Schecther} in order to compute the uncertainties on the SFRD. 

\begin{figure}
    \centering
    \includegraphics[width=\columnwidth]{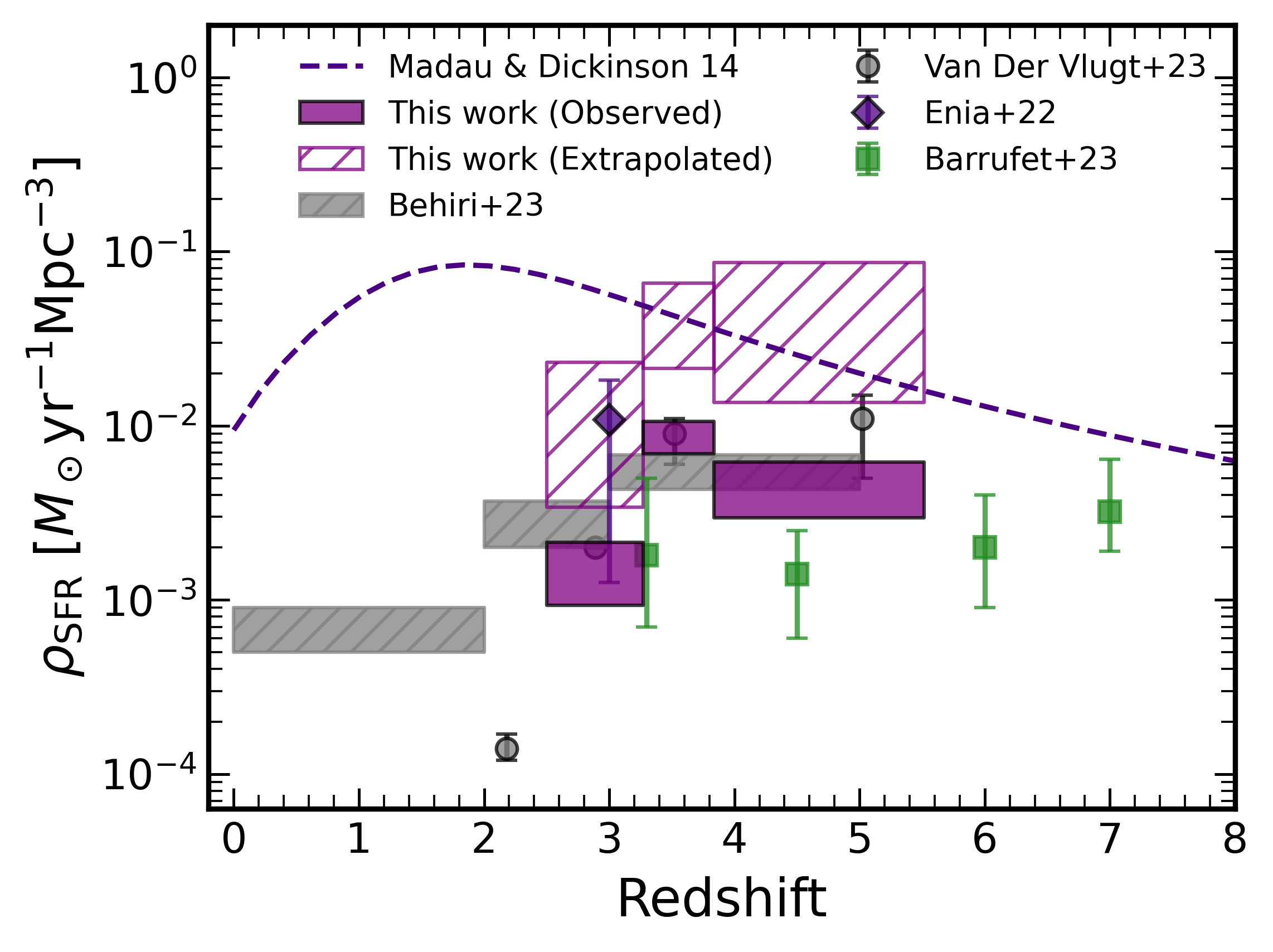}
    \caption{Cosmic Star Formation Rate Density. The values obtained by integrating the luminosity function of our RS-NIRfaint galaxies are reported in purple. The values obtained by integrating the observed range of luminosities are shown with full squares, while the empty boxes report the values obtained by integrating the LF on the full range of radio luminosities. The figure shows, for reference, the SFRD computed by \citet{madau_14} on ``optically-bright" galaxies (dashed indigo line), and those obtained by \citet{Behiri_23}, \citet{VanDerVlugt_23}, \citet{enia_22}, and \citet{Barrufet_23} (grey boxes, purple diamond, grey points, and green squares, respectively) for their ``NIRfaint galaxies" selected in the radio and with JWST.}
    \label{fig:SFRD}
\end{figure}

We perform two different integrals. The first one only includes the radio luminosities between the $L_{\rm min}$ related to the sensitivity of our radio survey and the maximum luminosity observed in each redshfit bin. The second extends over the full range of radio luminosities ($0\rightarrow+\infty$). While the first integral estimates the contribution of the galaxies that are actually observed in our surveys to the cosmic SFRD, the second one includes in this estimates the galaxies that are not observed in our survey but that we expect by extrapolating the radio LF at higher and lower luminosities. We underline that --- given these definitions --- the first value should be interpreted as a lower limit on the actual contribution of the RS-NIRfaint galaxies to the cosmic SFRD. The results of these integrals are summarized in \Tab\ref{tab:SFRD} and in \Fig\ref{fig:SFRD}. It is possible to notice how our estimates have large uncertainties. The reason for this can be found in the {\rev lack of strong constraints} on the $L_\star$ parameter of the LF. Being our galaxies unable to trace the faint end of the LF, our analysis can only pose a robust upper limit on the location of the ``knee". Deeper radio data would be needed to better constrain this value (and, consequently, the contribution to the SFRD).

\subsection{How much do the radio-selected NIRfaint galaxies contribute to the cosmic SFRD?}

The results obtained in \Sec\ref{sec:SFRD} indicate an increasing contribution of the RS-NIRfaint galaxies to the cosmic SFRD from $z\sim3$ to $z\sim3.5$ and then decreasing until $z\sim4.5$. This result agrees with previous studies focusing on NIRdark/faint radio sources at $z>2.5$. Our ``observed" results are consistent (even though moderately lower at $z\sim3$) with those obtained by \citet{Behiri_23}, analyzing the same kind of sources in the COSMOS field (even though with a slightly different selection, see \Sec\ref{sec:selection}). In that study, no extrapolation was performed outside the observed range of luminosities, therefore it is not surprising that our ``extrapolated" results are significantly higher. Similarly, our estimation of the SFRD at $z\sim3$ agrees well with the lower limit presented in \citet{enia_22} when no extrapolation is performed.

Our ``extrapolated" results can be compared with the upper limit provided by \citet{enia_22} (analyzing NIRdark galaxies in the GOODS-N field) and with the estimates by \citet{VanDerVlugt_23} (analyzing analogous sources in the deep COSMOS-XS survey). Both these studies estimated the radio LF and computed the contribution to the cosmic SFRD by integrating it in the full range of radio luminosities ($0\rightarrow\infty$). It is important to notice, however, that the three studies estimate the contribution to the total SFRD in different ways. The other two studies, indeed, focused on smaller fields compared with COSMOS-Web ($0.05$ deg$^2$ in GOODS-N and $0.1$ deg$^2$ in COSMOS-XS, instead of the $0.54$ deg$^2$ covered by our survey). This difference produced significantly smaller samples of NIRdark galaxies available for the analysis (9 sources for \citealt{enia_22} and 20 for \citealt{VanDerVlugt_23}, against the $\sim 120$ analysed here). To overcome this issue, \citet{enia_22} fixed three parameters in the Schechter fitting of the LF ($\sigma$, $\alpha$, and $L_\star$) to those obtained for the NIR-bright population, leaving only the normalization ($\Phi_\star$) as a free parameter of the fitting. Likewise, \citet{VanDerVlugt_23} computed the LF on the full sample of radio sources twice: once including the NIRdark galaxies and once excluding them: the contribution of these sources was then found by subtracting the two inferred SFRDs. 

Our results at $z\sim3.5$ agree well with the upper limit at the same redshift by \citet{enia_22}. Similarly, our result in the highest redshift bin ($z\sim4.5$) is compatible with the estimate by \citet{VanDerVlugt_23}. Their results are more in tension with ours at $z\sim3$ and $z\sim3.5$, with our estimates being up to one order of magnitude higher. A possible explanation of this discrepancy could reside in the accuracy of the photo-\textit{z} (our photometry includes the new NIRCam and MIRI photometry, that could result in more accurate redshift estimates; see e.g. \citealt{Barrufet_24}).

\subsection{Possible caveats of our analysis}
\label{sec:caveats}

Our analysis is not immune to possible biases. The most common one is an inaccuracy in the photometric redshifts estimated through SED fitting. Even with the new constraints obtained thanks to NIRCam and MIRI, the small number of photometric detections {\rev and the large number of upper limits make} our estimates uncertain. We reduced the possible impact of this issue on our results by employing in our analysis the full Gaussianized $p(z)$ given in output by \texttt{Cigale}, resulting in larger uncertainties on our LF and SFRD. This issue could be solved in the future with a spectroscopic follow-up of our galaxies (analogous to that performed by \citealt{Barrufet_24} with JWST, or those performed by \citealt{Jin_19,Chen_22,jin_22,Gentile_24b} with ALMA). 

Another caveat (more related to our radio selection) resides in the possible AGN contribution not unveiled by our $q_{\rm TIR}$ analysis. An overestimated radio luminosity due to star-formation could still bias our LF and SFR, producing a different behavior of the SFRD. This issue can be partially solved with a more complete photometric coverage at MIR wavelengths, where the thermal emission of a hot dusty torus surrounding the AGN should be visible (see e.g. \citealt{hickox_18} for a review and \citealt{Chien_24} for a recent application with MIRI data).

Finally, our analysis (especially the ``extrapolated" value of the SFRD) relies on a strong assumption about the shape of the LF. In our fitting procedure, we fixed the two parameters of the modified Schechter function $\alpha$ and $\sigma$ to those obtained by \citet{novak_17} for their sample of radio sources in the COSMOS field (i.e. the parent sample of our selection). This choice, however, assumes that the selection of more dust-obscured sources is not altering these parameters. This assumption could be incorrect, since we expect the number of NIRfaint sources to be lower in the low-SFR regime (e.g. for a lower production of dust from \textit{supernovae}) and higher in the high-SFR one (see e.g. \citealt{Whitaker_17} and A. Traina et al., \textit{in prep.}). This issue could be improved with a deeper radio survey (e.g. \citealt{VanDerVlugt_22}) giving us more information about the faint-end slope of the LF for our sources.

\begin{figure}
    \centering
    \includegraphics[width=\columnwidth]{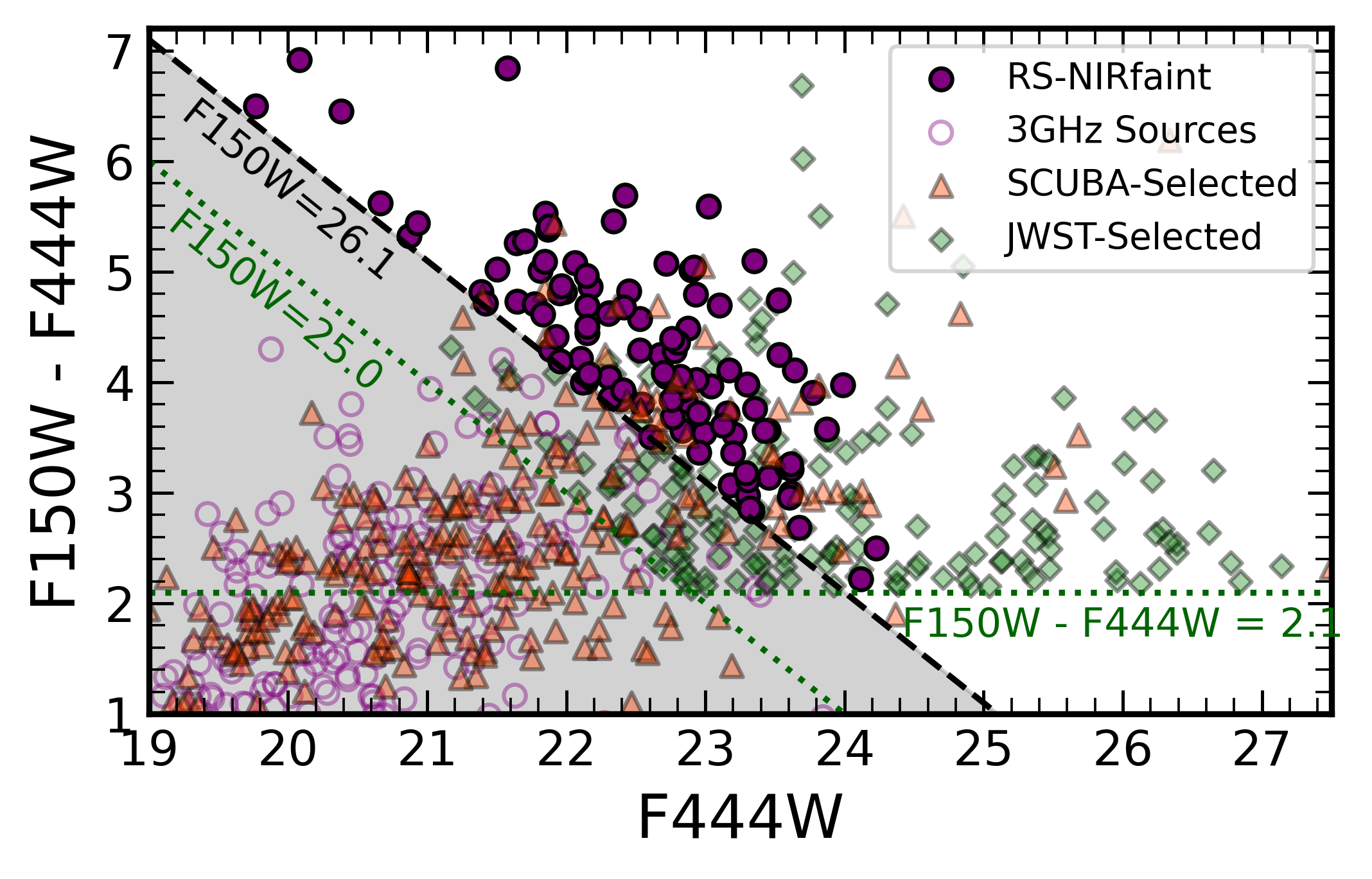}
    \caption{Our RS-NIRfaint galaxies in the F150W-F444W vs F444W color-magnitude plot. We report, for comparison, the JWST-selected DSFGs by \citet{Gottumukkala_23} and the SCUBA-selected ones by \citet{McKinney_24}. Even if we do not pose any constrain in the F150W-F444W color, we obtain that all our sources are above the threshold adopted by \citealt{Gottumukkala_23}, identifying our galaxies as a sub-population of the JWST-selected DSFGs. Similarly, part of the SCUBA-selected DSFGs satisfy our ``NIRfaint" criterion.}
    \label{fig:colors}
\end{figure}

\section{Discussion}
\label{sec:discussion}

\subsection{What is the effect of the selection for NIRfaint sources?}
\label{sec:other_pops}

To interpret the results presented in the previous sections, it is useful to compare our population of galaxies with other samples studied in the current literature. The first one is the total population of radio-detected sources, of which the RS-NIRfaint galaxies are a sub-population with $F150W>26.1$ mag. One of the main results of this comparison is shown in \Fig\ref{fig:photozs}, where we report the photo-\textit{z} estimated for our galaxies and those computed by M. Shuntov, L. Paquereau et al., (\textit{in prep.}) for the other 3 GHz sources in the COMSOS-Web survey with $F150W>26.1$ mag. It is immediately possible to see that our galaxies are --- on average --- located at higher redshifts with respect to the rest of the population. An analogous result has been shown by \citet{VanDerVlugt_23} for their NIRdark galaxies selected in the deep radio survey COSMOS-XS and by \citet{talia_21} and \citet{Behiri_23} in their previous analysis of RS-NIRdark galaxies in the COSMOS field. The same idea that NIRfaint DSFGs represent the high-redshift end of their parent distributions is also present in other studies selecting DSFGs at other wavelengths (see e.g. the studies on NIRfaint sub-millimeter galaxies by \citealt{Simpson_14,Franco_18,Smail_21,McKinney_24}). {\rev This result is likely connected to the existence of the $z_{\rm NIR}$ introduced in \Sec\ref{sec:LF} (see also \Fig\ref{fig:ZMinMax}) directly arising from the requirement of faintness at NIR wavelengths.}

The different distribution of the photo-\textit{z} also affects the estimated luminosity functions. Here, we consider those by \citet{novak_17}, \citet{enia_22}, and \citet{VanDerVlugt_22} focusing on NIR-bright galaxies. These estimates are reported for reference in \Fig\ref{fig:LF}, while the best-fitting parameters of the modified Schechter function found by \citet{enia_22} are reported in \Tab\ref{tab:LF}. It is possible to notice how the normalization of the luminosity function for our NIRfaint galaxies is on average $\sim1.5-2$ dex lower than what is observed for optically-bright sources. The luminosity of the knee, instead, seems to be higher for our sources, even though compatible with the values from \citet{enia_22} within the (large) uncertainties.

The first result is not unexpected, since we are dealing with a population of heavily dust-obscured sources less common than NIR-bright galaxies (see e.g. the comparison between the number of 3 GHz sources in COSMOS-Web and those in our sample in \Sec\ref{sec:data}). The second result is more interesting, since it suggests that the high dust obscuration could positively correlate with the radio luminosity (and --- therefore --- with the SFR). This result is not completely new, since several previous studies (e.g. \citealt{Whitaker_17} and Traina et al., \textit{in prep.}) highlighted how the dust-obscured star formation could dominate the high-SFR end of the star formation rate function, at least until the cosmic noon. The main consequence of the higher turn-off luminosity can be seen in \Fig\ref{fig:LF}, where is visible how the NIRfaint galaxies could dominate the LF in its bright end (even though, again, still compatible in most cases within the estimated uncertainties). The higher volume densities of our sources in the bright end also explains the large contribution to the SFRD when we integrate the LF on the whole range of radio luminosities to compute the SFRD (\Fig\ref{fig:SFRD}).

\subsection{What is the effect of the radio selection?}

Once confirmed that the radio selection is able to produce a sample of DSFGs, we want to analyse how this method compares with analogous strategies present in the current literature. In this section, we will focus on two main approaches: the selection by \citet{Barrufet_23} and \citet{Gottumukkala_23} based on the JWST colors, and that by \citet{McKinney_24} based on the (sub)mm detection.

\subsubsection{JWST-selected DSFGs}

We firstly compare our sources with those collected by \citet{Barrufet_23} and \citet{Gottumukkala_23} in the Cosmic Evolution Early Release Science (CEERS; \citealt{Finkelstein_23}). These studies mimic the DSFG selection initially performed by \citet{wang_19} with the so-called ``H-dropout" (i.e. galaxies selected through their H - [4.5] colors) by taking advantage of the new JWST photometry. More in detail, this selection couples the F150W-F444W$>2.1$ mag color cut with a magnitude cut in the F150W filter (F150W$>25.0$ mag in \citealt{Gottumukkala_23}). As visible in \Fig\ref{fig:colors}, our RS-NIRfaint galaxies represent a sub-sample of the sources analysed by \citet{Gottumukkala_23}, satisfying --- by construction --- the same criteria about faintness in F150W, having comparable F150W-F444W colors, but with the additional requirement of the radio detection. By comparing these samples, we can study how the radio selection affects the retrieved population of galaxies. Since the selection by \citet{Gottumukkala_23} includes fainter sources in the F150W filter, in the following we will focus on their sources satisfying our F150W$>$26.1 criterion.

The first notable difference is in the projected sky density of the two populations $6000\pm77$ deg$^{-2}$ for the sources by \citet{Gottumukkala_23} and $205\pm14$ deg$^{-2}$\footnote{For both estimates, we assume a Poissonian uncertainty $\sim\sqrt{N}$ on the observed number counts and do not correct for incompleteness} for our RS-NIRfaint galaxies. Then, by looking at the physical properties summarized in \Fig\ref{fig:properties}, we can see that the distributions of the photo-\textit{z} are quite similar until $z\sim4$, but with different behavior at higher redshifts. This result is not surprising given the much stronger \textit{k}-correction in the radio than in the F444W filter of NIRCam (see e.g. the shape of the SEDs in \Fig\ref{fig:SEDs}), biasing our sample towards lower-\textit{z} sources. 

Another notable difference consists in the reported values of SFR. As visible in \Fig\ref{fig:properties}, our RS-NIRfaint galaxies are $\sim1.5$ dex more star-forming than the sources analysed by \citet{Barrufet_23} and \citet{Gottumukkala_23}. On the one hand, this result is not surprising since the need for a radio detection excludes from our sample all the objects with low values of SFR. As noted by \citet{wang_19} and verified by \citet{Barrufet_24} with a spectroscopic follow-up of some JWST-selected objects, the red NIRCam colors alone are not enough to select DSFGs, since the same colors can be seen in quiescent galaxies. This ambiguity in the true nature of these sources can be solved with additional data from JWST (mainly spectra, as in \citealt{Barrufet_24}), or a more complete MIRI coverage (as in \citealt{perez-gonzalez_23}), ALMA (as in \citealt{wang_19}) or in the radio regime (as for our sources).

If the radio selection well explains why our sample lacks low-SFR galaxies, it does not explain why the high-SFR sources are not present in the sample by \citet{Gottumukkala_23}. A possible explanation resides in the photometric coverage of their sample. At the redshifts covered by their observations, no constraints on the rest-frame UV is available for the SED fitting, making quite uncertain the value of the SFR obtained through it. Similarly, we expect most of the star formation in these NIRfaint galaxies to be dust-obscured and --- therefore --- hard to constrain without some information on the FIR or radio emission, producing a likely underestimation of the SFR (see, e.g., some examples in \citealt{Xiao_23}).

The higher values of the SFR also explain one of the results visible in \Fig\ref{fig:SFRD}: even though the RS-NIRfaint galaxies are $\sim1.5$ dex less common than the JWST-selected sources, the higher values of SFR ($\sim 1.5$ dex) still produce compatible contribution to the cosmic SFRD for the ``observed" sources. The values obtained by extrapolating at higher and lower radio luminosities are obviously higher than those reported by \citet{Barrufet_23}, since their estimate only accounts for the detected sources.  

\subsubsection{SCUBA-selected DSFGs}

{\rev Together with the MIR- and radio-based selections, another} possibility to select DSFGs relies on the detection at (sub)mm wavelengths of the bright thermal emission by warm dust heated by ongoing star formation (see e.g. \citealt{casey_14} for a review). Galaxies selected through this procedure are commonly known as sub-millimeter galaxies (SMGs). Since this definition strongly relies on the depth of the observations employed to select these objects, in this study we only focus on the sources described in \citet{McKinney_24}, acknowledging that some differences might arise when other samples (detected in deeper or shallower surveys; see e.g. \citealt{daCunha_15,Dudzevicute_21}) are considered. These 289 galaxies were initially selected in the COSMOS field as sources with $S_{870 \mu {\rm m}}>2$ mJy in the SCUBA-2 observations performed during the S2COSMOS survey \citep{Simpson_19} and then followed-up with ALMA (unveiling in some cases the presence of multiple fainter sources contributing to the total flux of the SCUBA-2 objects). The SCUBA-Dive program described in \citet{McKinney_24} performs a deeper analysis of these sources taking advantage of the new JWST data coming from the COSMOS-Web survey.

Looking at the color-magnitude plot in \Fig\ref{fig:colors}, we can note only a partial overlap between the SCUBA-selected galaxies and our RS-NIRfaint ones, since most of the former are brighter (i.e. not NIRfaint) in the F150W filter. Since both samples are selected in the COSMOS field, we can cross-matching the two catalogs, finding only a partial overlap of 27 objects. We underline that the SCUBA-Dive programs only focuses on the SCUBA-2 sources with previous ALMA data, therefore some of our galaxies could still be included in that sample with additional data. To overcome this issue, we analyze the best-fitting SEDs computed with \texttt{Cigale} for our galaxies, obtaining that only 82 ($\sim 65$\%) of our sources would have a $S_{850 \mu m}>2$ mJy, being consistent with the initial cut employed on the SCUBA-2 maps by \citet{McKinney_24}.

By comparing the physical properties estimated through SED fitting (\Fig\ref{fig:properties}), we can see that the distributions of stellar masses and SFR are quite similar. More significant differences hold for the photo-\textit{z} (with our galaxies located --- on average --- at higher redshifts: $\langle z \rangle\sim 3.6$ for our sources and $\langle z \rangle\sim 2.6$ for those in SCUBA-Dive) and for the dust attenuation (with our sources being --- on average --- more dust-obscured; $\langle A_{\rm v} \rangle\sim 3.5$ mag against $\langle A_{\rm v} \rangle\sim 2.5$ mag). Both results are easily explained by the NIRfaint selection, biasing the sample towards more dust-obscured and higher-\textit{z} objects (see also the discussion in \Sec\ref{sec:other_pops}). 

A further confirmation of this result can be found in the comparison with the SCUBA-Dive sources satisfying the same magnitude cut as our RS-NIRfaint galaxies ($F150W>26.1$ mag). As visible from \Fig\ref{fig:properties}, these sources have analogous properties to our galaxies, indicating that the two selections are able to identify the same population of objects with different procedures.

\section{Summary}
\label{sec:summary}

In this paper, we presented the first analysis taking advantage of JWST data of radio-selected NIRfaint sources in the COSMOS field. These sources are defined as radio-detected sources with a counterpart at NIR wavelengths (unveiled with the deep NIRCam observations) fainter than the common depths reached by ground-based facilities. We obtained the following results:

\begin{itemize}
    \item The physical properties estimated through SED fitting indicate that the RS-NIRfaint galaxies are a population of highly dust-obscured ($\langle A_{\rm v} \rangle \sim3.5$ mag), massive ($\langle M_\star \rangle \sim10^{10.8}$ M$_\odot$) and star-forming galaxies ($\langle {\rm SFR} \rangle\sim300$ M$_\odot$ yr$^{-1}$) located at $\langle z \rangle\sim3.6$.
    \item Estimating the radio luminosity function of our sources and fitting it with a modified Schechter function, we find that its normalization ($\Phi_\star$) is $\sim1.5-$2 dex lower than that computed on radio-selected NIR-bright galaxies (e.g. \citealt{novak_17,enia_22,VanDerVlugt_22}), indicating that --- as expected --- DSFGs with faint optical/NIR counterparts are a rare population.
    \item Interestingly, the knee of the radio LF for our sources is brighter than for NIR-bright sources (even though compatible within the large estimated uncertainties). This result suggests that {\rev the fractional contribution of RS-NIRfaint sources (with respect to the overall population of galaxies at a given radio luminosity)} is negligible in the low-SFR end of the star formation rate function, while it becomes dominant in the high-SFR end, at least until $z\sim4.5$. This result confirms what has been found in previous studies focusing on the ratio between obscured and unobscured star formation in high-\textit{z} galaxies.
    \item By integrating the LF of our sources in the range of radio luminosities covered by our observations, we put a lower limit on their contribution to the cosmic SFRD. Our result shows an increasing contribution from $z\sim3$ to $z\sim3.5$ and then decreasing until $z\sim4.5$.
    \item By integrating the LF in the full range of radio luminosities ($0\rightarrow\infty$; i.e. extrapolating at higher and lower radio luminosities than what is actually observed), we increase the contribution to the total cosmic SFRD, reaching the same level of the NIR-bright galaxies analysed by \citet{madau_14}.
    \item When compared with the JWST selection of DSFGs carried out by \citet{Barrufet_23} and \citet{Gottumukkala_23}, we obtain that our radio selection generally misses the sources located at higher redshifts ($z>5.5$), as expected given the positive \textit{k}-correction affecting the radio emission. Moreover, our sources are $\sim 1.5$ dex more rare, and $\sim 1.5$ dex more star-forming. The main consequence of these differences, is that their contribution to the cosmic SFRD (not extrapolating at higher and lower radio luminosities) is compatible with that estimated by \citet{Barrufet_23} up to $z\sim5$.
    \item Finally, we compare our sources with the SCUBA-selected ones found in the COSMOS-Web survey by \citet{McKinney_24}, obtaining that our galaxies have analogous properties to their sources in terms of stellar mass and SFR. Our RS-NIRfaint galaxies are --- on average --- located at higher redshifts and more dust-obscured, as expected for our requirement about the faintness at NIR wavelengths. Comparing our sources with the SCUBA-selected galaxies satisfying the same $F150W>26.1$ mag requirement, we obtain more similar distributions of photo-\textit{z} and $A_{\rm v}$.
\end{itemize}

These results together justify the scientific interest in the population of RS-NIRfaint galaxies, picturing this selection as an efficient way of assembly statistically significant DSFGs in wide radio surveys.

\section*{Acknowledgments}
 We warmly thank the anonymous referee for his/her comments on the first version of this paper, which really allowed us to increase the overall quality of our study. F.G. thanks the department of astronomy of the University of Texas at Austin for the hospitality during the initial writing of this paper. F.G., M.T., and A. C. acknowledge the support from grant PRIN MIUR 2017-20173ML3WW\_001. ‘Opening the ALMA window on the cosmic evolution of gas, stars, and supermassive black holes’. MF acknowledges funding from the European Union’s Horizon 2020 research and innovation programme under the Marie Sklodowska-Curie grant agreement No 101148925. The french members of the COSMOS team acknowledge the support from CNES. The Cosmic Dawn Center (DAWN) is funded by the Danish National Research Foundation under grant DNRF140. M. V. acknowledges financial support from the Inter-University Institute for Data Intensive Astronomy (IDIA), a partnership of the University of Cape Town, the University of Pretoria and the University of the Western Cape, and from the South African Department of Science and Innovation's National Research Foundation under the ISARP RADIOMAP Joint Research Scheme (DSI-NRF Grant Number 150551) and the CPRR HIPPO Project (DSI-NRF Grant Number SRUG22031677). The Flatiron Institute is supported by the Simons Foundation.

\appendix

\section{\texttt{Cigale} parameters}
\label{App:SEDFitting}

In \Tab\ref{tab:cigale} we report the modules and parameters employed in our SED-fitting setup with \texttt{Cigale}.

\begin{table*}
\centering
\begin{threeparttable}
\renewcommand{\arraystretch}{1.1}
\caption{Parameters of the SED fitting performed with \texttt{Cigale}. A complete description of the modules and parameters can be found in \citet{boquien_19}.}
\label{tab:cigale}
\begin{tabular}{ccc}
\toprule
Parameter & Unit & Values\\
\midrule
\multicolumn{3}{c}{\texttt{sfhdelayed}}\\
\midrule
$\tau_{\rm main}$ & Myr & {\rev 600, 800,} 1000, 1800, 3000, 5000, 7000, 9000\\
Age$_{\rm main}$ & Myr & 1000, 3000, 5000, 8000, 11000, 12000\\ 
$\tau_{\rm burst}$ & Myr & 10 \\
Age$_{\rm burst}$ & Myr & 1, 10, 100, 300 \\
$f_{\rm burst}$ & - & {\rev  0, 0.05, 0.1, 0.15, 0.2}\\
Normalize & Bool & True\\ 
\midrule
\multicolumn{3}{c}{\texttt{bc03}}\\
\midrule
IMF & - & \citet{Chabrier_03}\\
$Z$ & - & 0.02 \\
Separation age & Myr & 10\\
\midrule
\multicolumn{3}{c}{\texttt{dustatt\_modified\_CF00}}\\
\midrule
$A_{\rm v}$ (ISM) & mag & 0.3, 0.9, 1.5, 2.1, 2.7, 3.3 \\
$\mu$ & - & 0.5, 0.8, 1.0 \\
$\alpha$ (ISM) & - & -0.7 \\
$\alpha$ (BC) & - & -0.7 \\
\midrule
\multicolumn{3}{c}{\texttt{dl2014}}\\
\midrule
$q_{\rm PAH}$ & - & 0.47, 1.12, 3.9\\
$U_{\rm min}$ & Habing & 5.0, 10.0, 25.0, 40, 50\\
$\alpha$ & - & 2.0\\
$\gamma$ & - & 0.0, 0.02\\
\midrule
\multicolumn{3}{c}{\texttt{nebular}}\\
\midrule
$\log(U)$ & - & -3.0 \\
$Z_{\rm gas}$ & - & 0.014 \\
$n_{\rm e}$ & - & 100.0 \\
$f_{\rm esc}$ & - & 0.0 \\
$f_{\rm dust}$ & - & 0.0 \\
lines width & km s$^{-1}$ & 300.0 \\
Emission & Bool & True\\
\midrule
\multicolumn{3}{c}{\texttt{redshifting}}\\
\midrule
$z$ & - & [0,8] with $\Delta z$=0.05\\
\bottomrule
\end{tabular}
\end{threeparttable}
\end{table*}

\bibliographystyle{aa}
\bibliography{Biblio} 

\end{document}